\documentclass[%
 amsmath,amssymb,
 aps,
]{revtex4-2}

\usepackage[all]{background} 
\usepackage{url}
\SetBgContents{Published by Physical Review Fluids}
\SetBgColor{gray}
\SetBgPosition{4.1,-12}
\SetBgOpacity{0.75}
\SetBgAngle{0.0}
\SetBgScale{2.0}

\usepackage{tikz,xcolor,hyperref}
\definecolor{lime}{HTML}{A6CE39}
\DeclareRobustCommand{\orcidicon}{%
	\begin{tikzpicture}
	\draw[lime, fill=lime] (0,0) 
	circle [radius=0.16] 
	node[white] {{\fontfamily{qag}\selectfont \tiny ID}};
	\draw[white, fill=white] (-0.0625,0.095) 
	circle [radius=0.007];
	\end{tikzpicture}
	\hspace{-2mm}
}
\foreach \x in {A, ..., Z}{%
	\expandafter\xdef\csname orcid\x\endcsname{\noexpand\href{https://orcid.org/\csname orcidauthor\x\endcsname}{\noexpand\orcidicon}}
}

\usepackage{graphicx}
\usepackage{dcolumn}
\usepackage{bm}

\newcommand{\bsub}{\begin{subequations}}
	\newcommand{\esub}{\end{subequations}}

\newcommand{\pt}{\mbox{\boldmath $\tau$}}

\newcommand{\pf}{\textbf{\emph{f}}}

\newcommand{\pn}{\textbf{\emph{n}}}

\newcommand{\pu}{\textbf{\emph{u}}}

\newcommand{\px}{\textbf{\emph{x}}}

\newcommand{\pat}{\partial}
\newcommand{\na}{\nabla}
\newcommand{\x}{\times}

\newcommand{\beq}{\begin{equation}}
\newcommand{\eeq}{\end{equation}}
\newcommand{\bsubeq}{\begin{subequations}}
	\newcommand{\esubeq}{\end{subequations}}
\newcommand{\beqn}{\begin{eqnarray}}
\newcommand{\eeqn}{\end{eqnarray}}

\newcommand{\lb}{\label}
\newcommand{\er}{\eqref}

\usepackage{hyperref}
\hypersetup{colorlinks,allcolors=blue}
\usepackage{amsmath}
\usepackage{multirow}
\usepackage{amsfonts,amssymb}
\usepackage{graphicx}
\usepackage{subfigure}
\usepackage[rel]{overpic}

\graphicspath{{figures/}}

\begin{document}

\preprint{APS/123-QED}

\title{Enhanced wind-farm performance using windbreaks}

\author{Luoqin Liu\orcidA{}}
 \email{luoqin.liu@utwente.nl}


\author{Richard J. A. M. Stevens\orcidC{}}
 \email{r.j.a.m.stevens@utwente.nl}

\affiliation{Physics of Fluids Group, Max Planck Center Twente for Complex Fluid Dynamics, University of Twente, 7500 AE Enschede, The Netherlands}%

\date{\today}

\begin{abstract}
The flow speed-up generated by windbreaks can be used to increase the power production of wind turbines. However, due to the increased drag imposed by the windbreaks, their use in large wind turbine arrays has been questioned. We use large eddy simulations to show that windbreaks can increase the power production of large wind farms. A crucial finding is that windbreaks in a wind farm should be much lower than for a single turbine case. In fact, the optimal windbreak for an isolated turbine can reduce wind farm performance. The optimal windbreak height in a wind farm namely depends on the right balance between flow speed-up over the windbreak and the drag imposed by all windbreaks in the farm. The increased performance is a result of the favorable total pressure flux created by the windbreaks.
\end{abstract}

\maketitle

\section{Introduction}\lb{sec.introduction}

A wind farm is an assembly of wind turbines that converts wind energy into electricity. The wind turbines extract kinetic energy from the flow, which leads to the formation of wakes. Further downwind, the wind turbine wakes expand and interact with the atmospheric boundary layer (ABL). In large wind farms the performance of downwind turbines is significantly impacted by these wind turbine wakes \citep{ste17, men19, por20}. Therefore, wind farm layout optimization \citep{kus10, sha16c, ste17b} and control methods \citep{goi15, how19, bas19} to improve the performance of large wind farms are actively investigated.

Windbreaks have been used for centuries to reduce near-surface wind speeds and soil erosion, control snowdrift, and provide a favorable micro-climate for humans, animals, and plants \citep{cou74, law78, bra83, fin83, wil85, fan97, wan01, rau01, don07, bou08, tob18, bas19b}. When the windbreak porosity is sufficiently low, the flow behind and above the windbreak can be divided into three distinct regions \citep{cou74, tob17a}. According to the definitions of \citet{cou74}, the region downwind of the windbreak and adjacent to the wall is the wall region, where the velocity follows the logarithmic law. The region downwind of the windbreak and above the wall region is the mixing region, where the perturbation velocity allows a self-preserving solution. The region above the windbreak and the mixed region is the external region, an inviscid perturbation of the boundary layer flow. The streamwise flow velocity is reduced in the wall and mixing regions due to windbreak wake formation. However, in the external region, the flow velocity increases due to the flow speed-up over the windbreak. In addition to the porosity, the windbreak wake also depends on the windbreak aspect ratio, the incidence angle, and the surface roughness \citep{law78}. Recently, \citet{tob18} found experimentally that the so-called wake-moment coefficient does not change when the windbreak aspect ratio is larger than 10; however, it may become lower for an aspect ratio of 5. \citet{bas19b} showed in simulations that the main features of the wake flow are the formation of a bottom-attached recirculation region behind the windbreak and corotating vortices originating from eddies shed inside the separated shear layer at the top edge of the windbreak.

Windbreaks can also be used to increase the power production of wind turbines due to the local speed-up effect over the windbreaks. \citet{tob17a} showed using inviscid flow theory that, for low windbreaks, the power production increases approximately linearly with windbreak height. Their wind-tunnel measurements confirmed this result. \citet{kim19} analyzed the hourly Supervisory Control and Data Acquisition (SCADA) data of the Gunsan wind farm, which consists of a row of $10$ wind turbines spaced about four turbine diameters apart. The SCADA data showed that the turbine power production is increased by about $1.5\%$ due to the flow speed-up over a nearby seawall, which was in agreement with their Reynolds-averaged Navier-Stokes simulation model predictions.

\citet{tob17b} investigated the effect of windbreaks on the performance of infinite wind farms theoretically and numerically. Following the idea of the top-down model pioneered by \citet{fra92} and further developed by \citet{cal10}, the windbreaks were parametrized as increased surface roughness. These model calculations showed that windbreaks in an infinite wind farm reduce the wind speed at hub-height due to the increased drag. Their large-eddy simulations (LES) confirmed these model calculations, which suggests that windbreaks are not effective in improving the performance of large wind farms. \citet{zha18a} carried out wind tunnel measurements to quantify the effects of windbreaks in very large wind farms. Their model wind farm consists of 5 columns and 20 rows, and their measurements show that the negative impact of the windbreak wakes outweighs the local speed-up effect over the windbreaks.

While previous studies showed that windbreaks can increase the power production of isolated turbines, it is argued that the additional drag imposed by the windbreaks makes their use ineffective in wind farms. In the present study, we use LES to explore whether windbreaks can improve wind farm power production. In section~\ref{sec.LES} we discuss the simulation method and considered wind farm configuration. In section~\ref{sec.results} we will show that windbreak can be used to improve wind farm performance and analyze the effect of windbreaks on the flow using a kinetic energy budget analysis. The conclusions will be presented in section ~\ref{sec.conclusions}.

\section{Numerical method and validation}\label{sec.LES}

\subsection{Numerical method}

The simulations are performed with a code that solves the spatially-filtered continuity and momentum equations, where the subgrid-scale (SGS) shear stress is parameterized using the scale-dependent Lagrangian dynamic model \citep{bou05}. 
The computational grid is uniform in the horizontal and vertical directions and staggered in the wall-normal direction. The first vertical velocity grid plane is located at the ground, while the first horizontal velocities grid plane is located at half a vertical grid spacing above the ground. We use a pseudo-spectral discretization, and thus periodic boundary conditions, in the horizontal directions and a second-order finite difference method in the vertical direction. We enforce a zero vertical velocity at the top boundary and a zero shear stress. At the bottom boundary, we employ the classic wall model to determine the wall SGS stress \citep{moe84, bou05}. Time integration is performed using a second-order Adams--Bashforth method, and the projection method is used to ensure that the velocity field is divergence-free. We use the concurrent precursor method to generate turbulent inflow conditions that match atmospheric turbulence \citep{ste14}. For a detailed description and validation of our code we refer the reader to Refs.\ \citep{gad20, liu20, liu20d}

The wind turbines are modeled using an actuator disk approach in which the free-stream velocity $U_\infty$ is used to calculate the turbine force $F_{\rm wt}$ 
\begin{equation}
F_{\rm wt} = - \frac{1}{2}\rho C_T U_\infty^2 \frac{\pi}{4} D^2, 
\end{equation}
where $\rho$ is the density of the fluid, $C_T$ is the thrust coefficient based on $U_\infty$, and $D$ is the turbine diameter. However, when a turbine operates in the wake of upwind turbines or windbreaks, the free-stream velocity is not readily available. \citet{cal10} pointed out that actuator disk theory can be used to model the turbine force as follows
\begin{equation}
F_{\rm wt} = - \frac{1}{2}\rho C_T' U_d^2 \frac{\pi}{4} D^2, 
\end{equation}
where $C_T'$ is the thrust coefficient based on the disk-averaged velocity $U_d$. From the momentum theory (see, e.g. \citep{bur01}), it follows that these two thrust coefficients are related by $C_T'=C_T/(1-a)^2$, where $a$ is the axial induction factor. 

Following \citet{tob17b} the windbreaks are modeled in a similar way by defining the windbreak force $F_{\rm wb}$ as
\beq \lb{eq.Fwb}
F_{\rm wb} = -\frac{1}{2} \rho k U_w^2 bh,
\eeq 
where $k$ is the pressure coefficient, $U_w$ is the windbreak-averaged velocity, $b$ is the windbreak width, and $h$ is the windbreak height. \citet{tay44} found that for sparse screens $k$ can be related to the porosity $\eta$ by the following empirical relationship:
\beq\lb{eq.Cp}
k = c_e \frac{1-\eta}{\eta^2},
\eeq 
where $c_e=O(1)$ is the drag coefficient of a screen element. Equation~\er{eq.Cp} is based on velocity enhancement due to flow constriction through hole and may not be accurate for dense-screens or vegetative barrier. The determination of more general relationship between $k$ and $\eta$ is out of the scope of this study, for which we refer the reader to Refs.~\citep{rau01, ste18-jfm}. In this paper, we use Eq.~\er{eq.Cp} and document the properties of the windbreak by giving $\eta$. In particular, we use $c_e=1.2$ such that the simulation results agree well with field measurements (see below in \S~\ref{sec.validation}).

\subsection{Numerical validation}\lb{sec.validation}

To validate our simulation approach for the windbreaks we compare our simulation results with the classical field measurements for flow over windbreaks by \citet{fin83}. The roughness height at the site was $z_0 = 2.0$~mm, the windbreak height was $h=1.2$~m, and the windbreak porosity is $\eta = 0.5$. To simulate this case we consider a computational domain of $L_x \times L_y \times L_z = 96h \times 12h \times 8h$ in the streamwise, spanwise, and vertical directions, respectively, which is discretized on a $N_x \times N_y \times N_z =384 \times 48 \times 129$ grid.

\begin{figure} 
\centering
\begin{overpic}[width=0.5\textwidth]{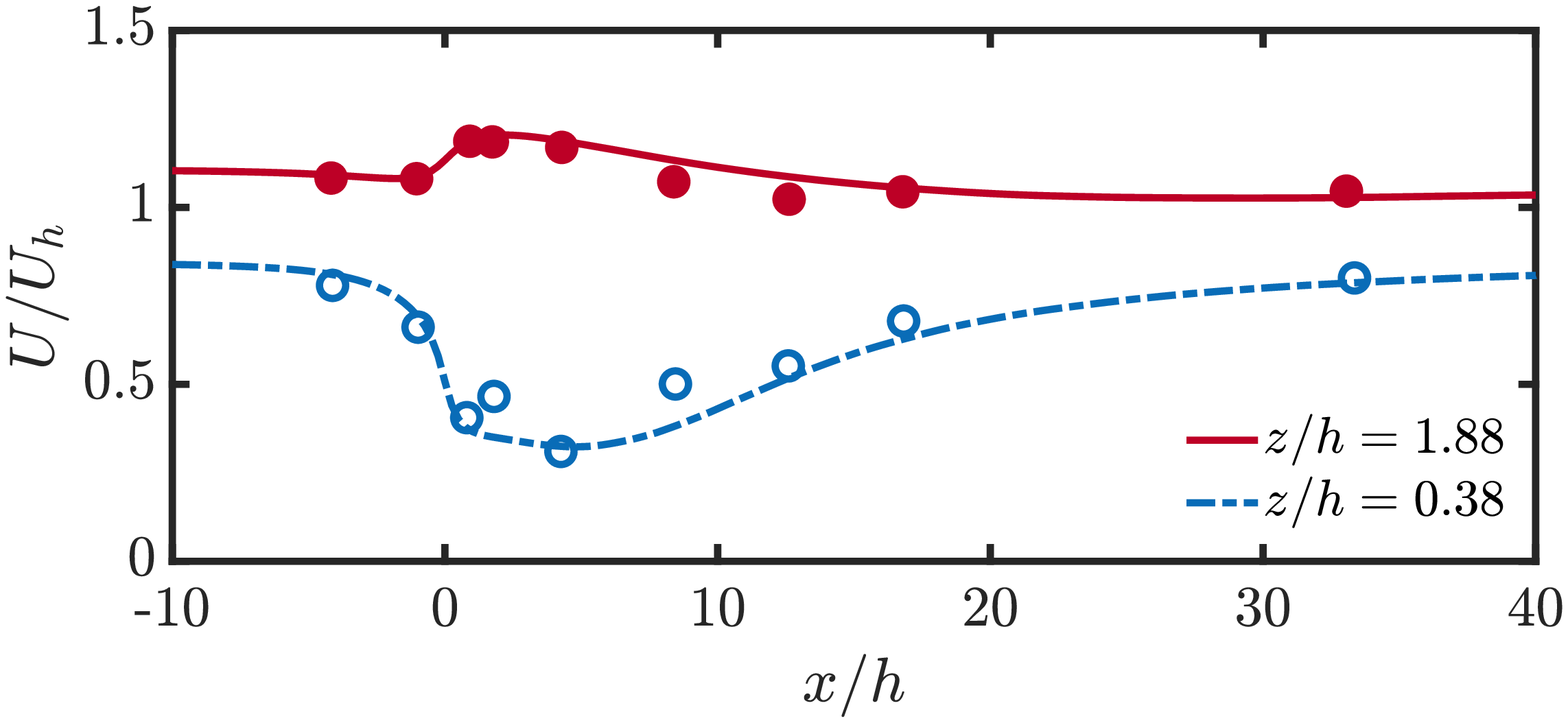} 
\put(0,40.8){$(a)$}
\end{overpic} 
\hspace{2mm}
\begin{overpic}[width=0.23\textwidth]{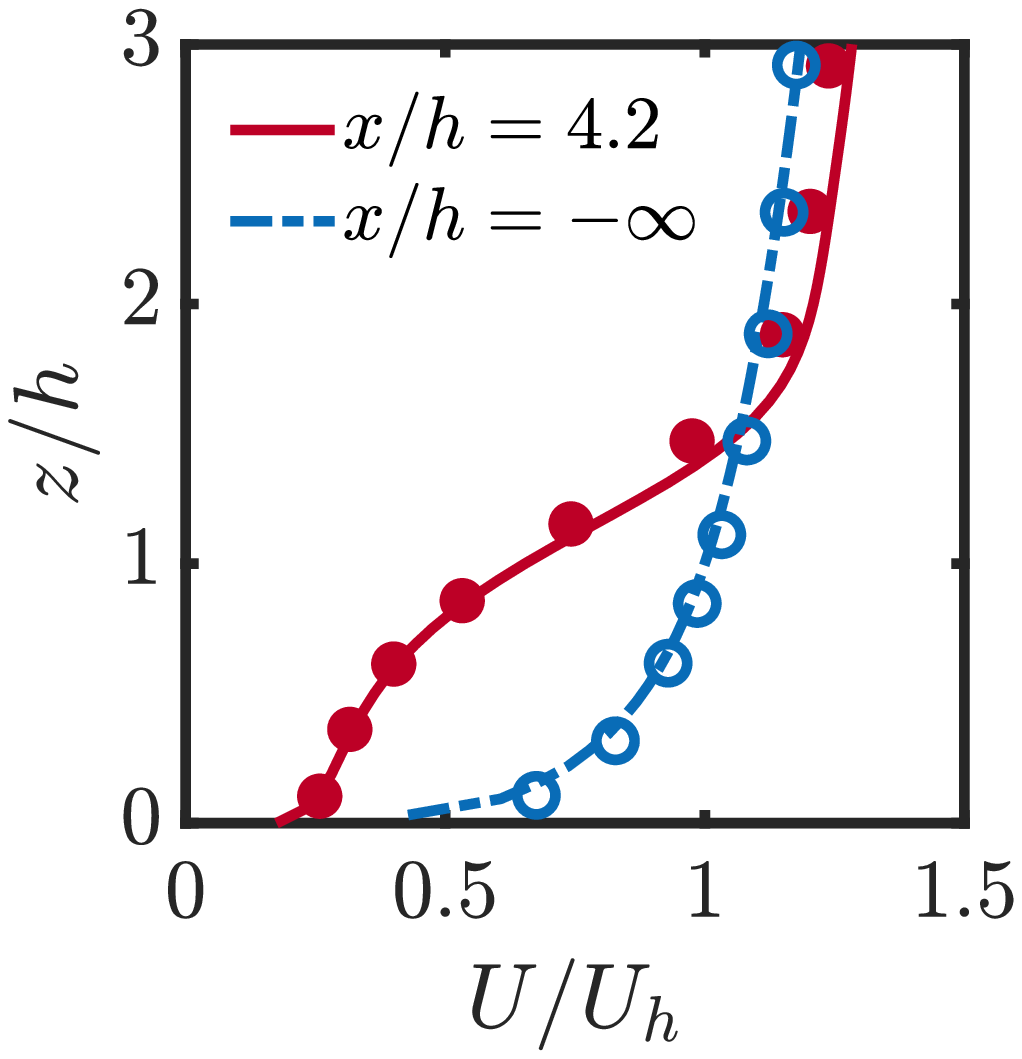} 
\put(0,89){$(b)$}
\end{overpic}
\begin{overpic}[width=0.7\textwidth]{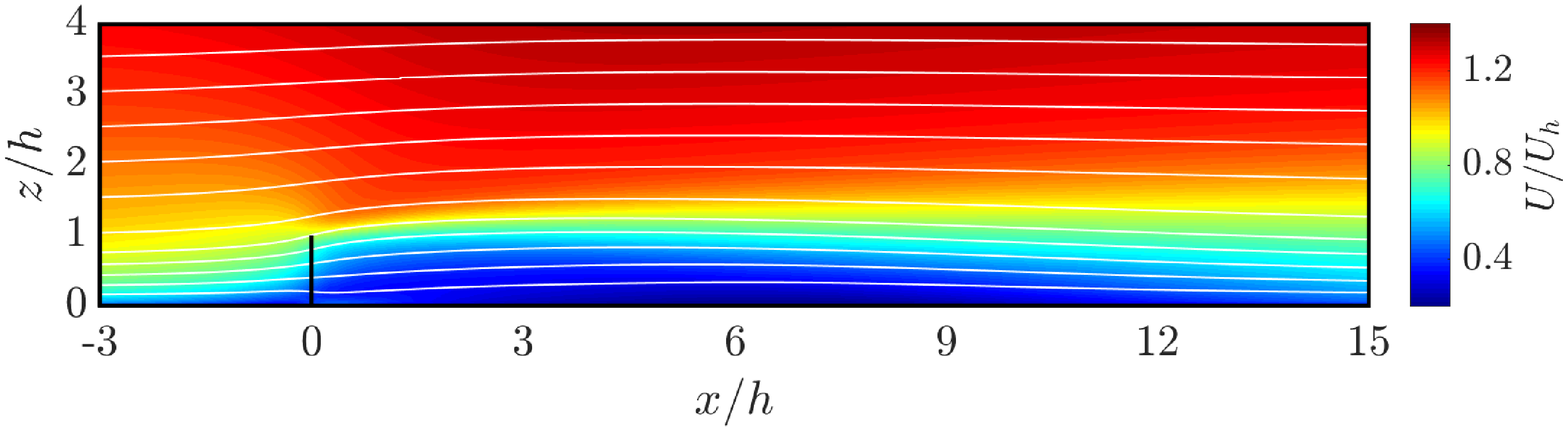} 
\put(0,24.5){$(c)$}
\end{overpic} 
\caption{(a) Horizontal and (b) vertical profiles of the normalized streamwise velocity $U/U_h$. Lines: LES data; symbols: field data taken from \citet{fin83} as presented in the reanalysis by \citet{wil85}. (c) Streamline pattern for flow through a windbreak with porosity $\eta=0.5$. The background colour indicates $U/U_h$ and the solid line denotes the windbreak.}
\label{fig.validation}
\end{figure}

Figure~\ref{fig.validation}(a,b) shows that the time-averaged horizontal and vertical profiles of the normalized streamwise velocity $U/U_h$, where $U$ is the streamwise velocity and $U_h$ is the velocity at windbreak height far upwind, is in good agreement with the field measurement data by \citet{fin83}. The figure shows that our simulation method correctly captures the speed-up over the windbreak and the subsequent flow recovery further downwind. Figure~\ref{fig.validation}(c) shows the visualization of the normalized streamwise velocity and the corresponding streamlines for this case, which reveals that, due to its high porosity, no recirculation zone is formed behind this windbreak.

\begin{figure} [!tb] 
\centering
\begin{overpic}[width=0.7\textwidth]{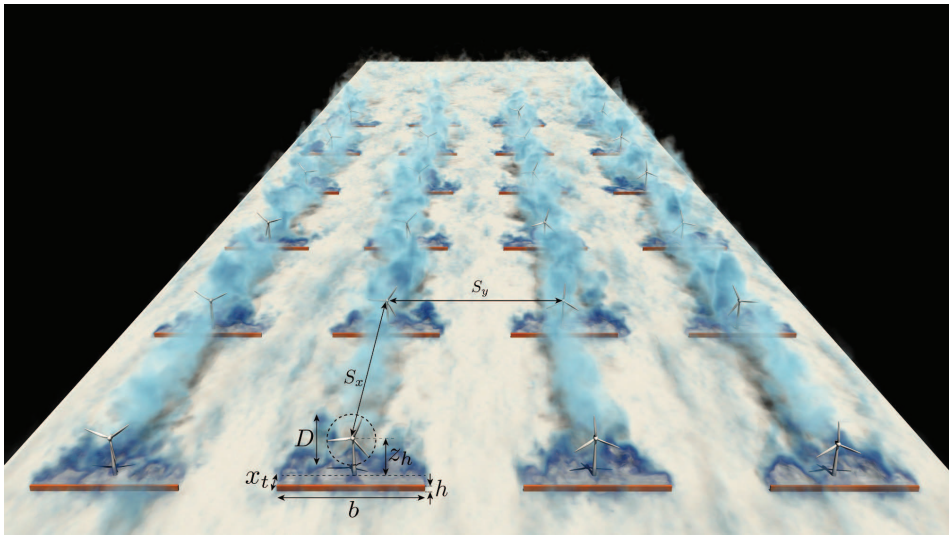} 
\end{overpic} 
\caption{Schematic of the layout of a wind farm with windbreaks. In this study, the streamwise and spanwise turbine spacings are $S_x=8D$ and $S_y=5D$, and the values of windbreak width $b$, windbreak height $h$, and the spacing between the windbreak and turbine $x_t$ are varied. The effects of windbreak porosity and the number of windbreaks are also studied (not shown in the figure). The background color presents the instantaneous streamwise velocity distribution from case 39 (see Table~\ref{tab1}). The blue color levels are between $[-1, 18]$ such that dark blue denotes the windbreak wake and light blue denotes the turbine wake. A video showing the development of the flow field is provided in the Supplemental Material \citep{liu21-prf}. } 
\label{fig.sketch}
\end{figure}

\begin{table} [!tb]
\begin{center}
\caption{Summary of the performed simulations. Case $1$ is the windbreak simulation presented in Fig.~\ref{fig.linear-law}. Case 2 and 37 are reference wind farm simulations without windbreaks and in all other simulations various windbreak configurations are considered. Here, $L_x \times L_y \times L_z$ is the domain size, $N_x \times N_y \times N_z$ is the grid resolution, $N_{\rm wb}$ is the number of windbreaks, $b$ and $h$ are the width and height of windbreak, $D$ is the turbine diameter, $\eta$ is the windbreak porosity, and $x_t$ is the distance between the windbreak and wind turbine, see Fig.~\ref{fig.sketch}. $P_{\rm t}/P_{\rm r}$ is the ratio of the wind farm power production with windbreaks ($P_{\rm t}$) to the corresponding wind farm without windbreaks ($P_{\rm r}$). $P_1/P_0$ is the ratio of the power production of the first row of the farm with windbreaks ($P_{1}$) to the corresponding reference without windbreak ($P_{0}$). }
\label{tab1}
\begin{tabular}{cccccccccc}
\hline 
\hline 
Case & $L_x \x L_y \x L_z$ & $N_x \x N_y \x N_z$ & $N_{\rm wb}$ & $h/D$ & $b/D$ & $\eta$ & $x_t/h$ & $P_{\rm t}/P_{\rm r}$ & $P_1/P_0$ \\
\hline 
1 & $40D \times 5D \times 5D$ & $384\x 48 \x 145$ & 1 & 0.24 & 5 & 0.03 & $-$ & $-$ & $-$ \\ 
2 & $80D \times 5D \times 5D$ & $768\x 48 \x 145$ & 0 & 0 & 0 & 1 & $-$ & 1 & 1 \\ 
3 & $80D \times 5D \times 5D$ & $768\x 48 \x 289$ & 6 & 0.12 & 5 & 0.03 & 2 & 1.11 & 1.24 \\
4 & $80D \times 5D \times 5D$ & $768\x 48 \x 289$ & 6 & 0.12 & 5 & 0.03 & 4 & 1.13 & 1.26 \\
5 & $80D \times 5D \times 5D$ & $768\x 48 \x 289$ & 6 & 0.12 & 5 & 0.03 & 6 & 1.14 & 1.30 \\
6 & $80D \times 5D \times 5D$ & $768\x 48 \x 289$ & 6 & 0.12 & 5 & 0.03 & 8 & 1.10 & 1.24 \\
7 & $80D \times 5D \times 5D$ & $768\x 48 \x 145$ & 6 & 0.24 & 5 & 0.03 & 2 & 1.25 & 1.51 \\
8 & $80D \times 5D \times 5D$ & $768\x 48 \x 145$ & 6 & 0.24 & 5 & 0.03 & 4 & 1.16 & 1.45 \\
9 & $80D \times 5D \times 5D$ & $768\x 48 \x 145$ & 6 & 0.24 & 5 & 0.03 & 6 & 1.03 & 1.36 \\
10 & $80D \times 5D \times 5D$ & $768\x 48 \x 145$ & 6 & 0.24 & 5 & 0.03 & 8 & 0.93 & 1.24 \\
11 & $80D \times 5D \times 5D$ & $768\x 48 \x 145$ & 6 & 0.36 & 5 & 0.03 & 2 & 1.13 & 1.59 \\
12 & $80D \times 5D \times 5D$ & $768\x 48 \x 145$ & 6 & 0.36 & 5 & 0.03 & 4 & 0.98 & 1.46 \\
13 & $80D \times 5D \times 5D$ & $768\x 48 \x 145$ & 6 & 0.36 & 5 & 0.03 & 6 & 0.80 & 1.24 \\
14 & $80D \times 5D \times 5D$ & $768\x 48 \x 145$ & 6 & 0.36 & 5 & 0.03 & 8 & 0.71 & 0.99 \\
15 & $80D \times 5D \times 5D$ & $768\x 48 \x 145$ & 6 & 0.48 & 5 & 0.03 & 2 & 0.83 & 0.97 \\
16 & $80D \times 5D \times 5D$ & $768\x 48 \x 145$ & 6 & 0.48 & 5 & 0.03 & 4 & 0.67 & 0.76 \\
17 & $80D \times 5D \times 5D$ & $768\x 48 \x 145$ & 6 & 0.48 & 5 & 0.03 & 6 & 0.56 & 0.65 \\
18 & $80D \times 5D \times 5D$ & $768\x 48 \x 145$ & 6 & 0.48 & 5 & 0.03 & 8 & 0.51 & 0.56 \\
19 & $80D \times 5D \times 5D$ & $768\x 48 \x 289$ & 1 & 0.12 & 5 & 0.03 & 6 & 1.04 & 1.24 \\
20 & $80D \times 5D \times 5D$ & $768\x 48 \x 289$ & 2 & 0.12 & 5 & 0.03 & 6 & 1.05 & 1.26 \\
21 & $80D \times 5D \times 5D$ & $768\x 48 \x 289$ & 3 & 0.12 & 5 & 0.03 & 6 & 1.06 & 1.27 \\
22 & $80D \times 5D \times 5D$ & $768\x 48 \x 145$ & 1 & 0.24 & 5 & 0.03 & 4 & 1.09 & 1.46 \\
23 & $80D \times 5D \times 5D$ & $768\x 48 \x 145$ & 2 & 0.24 & 5 & 0.03 & 4 & 1.09 & 1.49 \\
24 & $80D \times 5D \times 5D$ & $768\x 48 \x 145$ & 3 & 0.24 & 5 & 0.03 & 4 & 1.10 & 1.49 \\
25 & $80D \times 5D \times 5D$ & $768\x 48 \x 289$ & 6 & 0.12 & 2 & 0.03 & 6 & 1.07 & 1.13 \\
26 & $80D \times 5D \times 5D$ & $768\x 48 \x 289$ & 6 & 0.12 & 3 & 0.03 & 6 & 1.14 & 1.28 \\
27 & $80D \times 5D \times 5D$ & $768\x 48 \x 289$ & 6 & 0.12 & 4 & 0.03 & 6 & 1.13 & 1.26 \\
28 & $80D \times 5D \times 5D$ & $768\x 48 \x 145$ & 6 & 0.24 & 2 & 0.03 & 2 & 1.10 & 1.21 \\
29 & $80D \times 5D \times 5D$ & $768\x 48 \x 145$ & 6 & 0.24 & 3 & 0.03 & 2 & 1.22 & 1.47 \\
30 & $80D \times 5D \times 5D$ & $768\x 48 \x 145$ & 6 & 0.24 & 4 & 0.03 & 2 & 1.22 & 1.47 \\
31 & $80D \times 5D \times 5D$ & $768\x 48 \x 289$ & 6 & 0.12 & 5 & 0.15 & 6 & 1.10 & 1.22 \\
32 & $80D \times 5D \times 5D$ & $768\x 48 \x 289$ & 6 & 0.12 & 5 & 0.23 & 6 & 1.07 & 1.20 \\
33 & $80D \times 5D \times 5D$ & $768\x 48 \x 289$ & 6 & 0.12 & 5 & 0.32 & 6 & 1.06 & 1.16 \\
34 & $80D \times 5D \times 5D$ & $768\x 48 \x 145$ & 6 & 0.24 & 5 & 0.15 & 2 & 1.10 & 1.36 \\
35 & $80D \times 5D \times 5D$ & $768\x 48 \x 145$ & 6 & 0.24 & 5 & 0.23 & 2 & 1.06 & 1.28 \\
36 & $80D \times 5D \times 5D$ & $768\x 48 \x 145$ & 6 & 0.24 & 5 & 0.32 & 2 & 1.03 & 1.24 \\
37 & $80D \times 20D \times 5D$ & $768\x 192 \x 161$ & 0 & 0 & 0 & 1 & $-$ & 1 & 1 \\
38 & $80D \times 20D \times 5D$ & $768\x 192 \x 321$ & 6 & 0.12 & 20 & 0.03 & 5 & 1.14 & 1.28 \\
39 & $80D \times 20D \times 5D$ & $768\x 192 \x 161$ & 6 & 0.24 & 20 & 0.03 & 2 & 1.22 & 1.45 \\
\hline 
\hline 
\end{tabular}
\end{center}
\end{table}

\subsection{Considered cases}\lb{sec.cases}

To study whether windbreaks can increase the power production of a wind farm, we perform a series of LES of wind farms with and without windbreaks. \citet{tob17b} showed that windbreaks reduce the power production of infinite wind farms. Here we consider a wind farm with six rows in the downwind direction. Following the work of \citet{tob17b} the flow is driven by a constant pressure gradient $u_*^2/\delta = 5 \times 10^{-4}$~m/s$^2$, where the friction velocity $u_*=0.5$~m/s and the ABL thickness $\delta = 500$~m. The roughness height is $z_0=0.01$~m. The turbine diameter $D$ and hub-height $z_h$ are $100$~m, and the turbine thrust coefficient is $C_T'=0.9$. The streamwise and spanwise turbine spacings are $S_x=8D$ and $S_y=5D$, respectively. A sketch of the considered wind farm layout with windbreaks is shown in Fig.~\ref{fig.sketch}, which illustrates the definition of all geometrical quantities considered in this study. A summary of the performed simulations is given in Table~\ref{tab1}. All simulations are performed for 54 flow-through times, and statistics are averaged over the last 11 flow-through times to ensure that statistically converged data are obtained.

The first row of wind turbines is always located at $8D$ downwind of the entrance. A long fringe region of 10\% of the computational section is used to ensure a smooth transition from the flow formed behind the wind farm towards the applied inflow condition \citep{ste14}. For all wind farm simulations, the streamwise domain size is $L_x=80D$, and the vertical domain size is $L_z=5D$. To save computational resources, the spanwise domain size of most of the simulations is $L_y=5D$. We verified that this choice does not affect the main results by performing additional simulations in a wider spanwise domain of $L_y=20D$. Table~\ref{tab1} shows that the results obtained in a $L_y=20D$ and $L_y=5D$ domain agree very well, see the results for low $h/D=0.12$ (cases 4, 5, and 38) and intermediate windbreak heights $h/D=0.24$ (cases 7 and 39).

\section{Results}\label{sec.results}

\subsection{Flow over windbreak}\lb{sec.windbreak}

As discussed before, the speed-up over the windbreak can increase the power production of turbines downwind of the windbreak. A first-order estimate gives that the expected power production increase can be obtained from the flow field over a windbreak. \citet{tob17a} found that, for low windbreaks, there is an approximately linear relationship between the power increase $\Delta P$ and the height ratio between the windbreak and hub-height $h/z_h$, 
\beq\lb{eq.P-P0}
\frac{\Delta P}{P_0} = \left( 1 + \frac{\Delta U}{U_{\rm hub}} \right)^3 -1 = \alpha \frac{h}{z_h} + \beta,
\eeq
where $P_0$ is the power production of a stand-alone wind turbine, $U_{\rm hub}$ is the incoming streamwise velocity at hub-height, $\Delta U$ is the increase in the streamwise velocity at hub-height caused by the windbreak, and $\alpha=\alpha(\eta, x_t)$ and $\beta=\beta(\eta, x_t)$ are fitted constants. We set $\beta=0$ as there should be no power increase when there is no windbreak ($h/z_h=0$). Experimental data show that $\alpha$ increases with decreasing $\eta$ because the speed-up over the windbreak is stronger for lower porosity \citep{fan97, don07, tob17a}.

\begin{figure}  [!tb]
\centering
\begin{overpic}[width=0.8\textwidth]{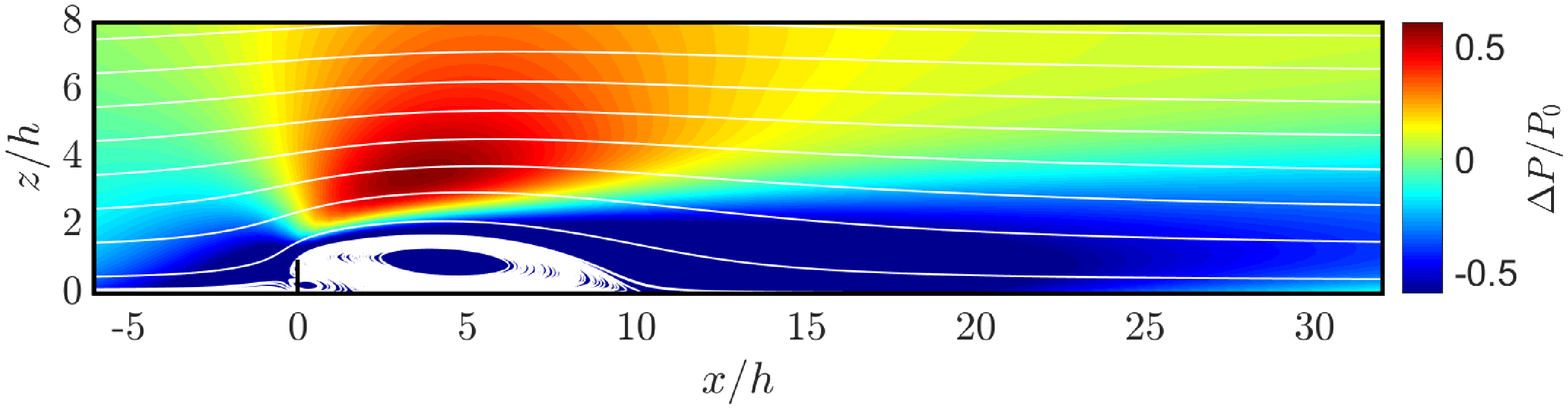} 
\put(0,22.5){$(a)$}
\end{overpic} \\
\begin{overpic}[width=0.7\textwidth]{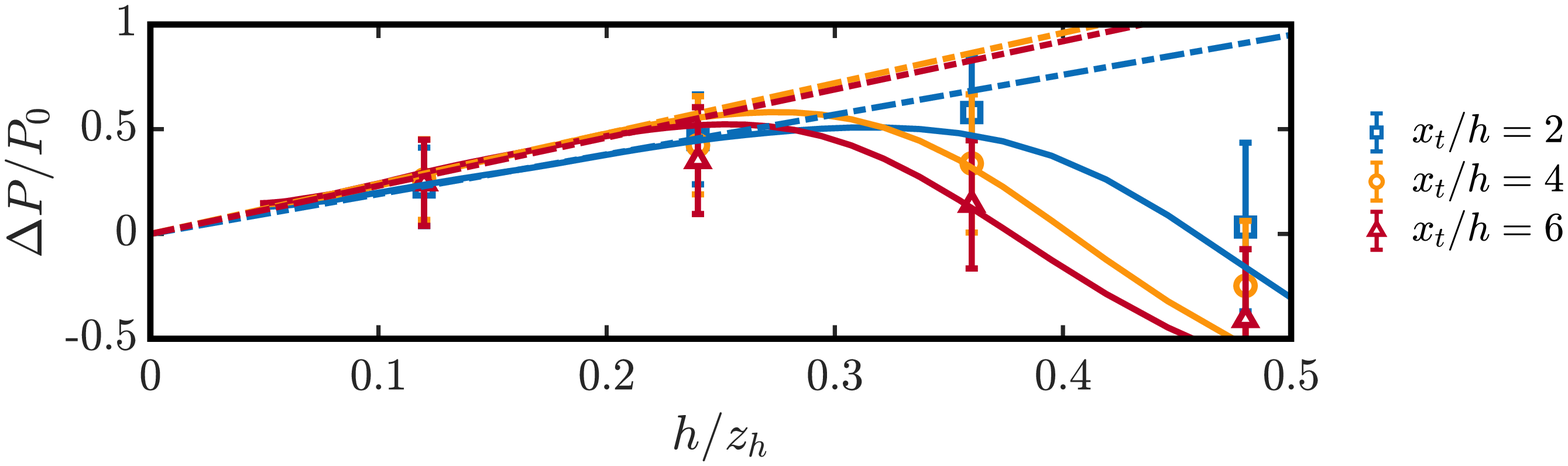} 
\put(0,25.5){$(b)$}
\end{overpic} 
\caption{(a) Streamline pattern for flow through a windbreak with porosity $\eta=0.03$. The background colour shows the estimated power increase according to the first equality in Eq.~\er{eq.P-P0}. The solid line denotes the windbreak. (b) Power production increase for a turbine located at $x_t/h=2$ (blue), $x_t/h=4$ (yellow), and $x_t/h=6$ (red). The solid lines indicate the predicted power production increase calculated from the flow field over a windbreak obtained from LES, see panel (a). The dashed lines are fits to solid lines using the second equality in Eq.~\er{eq.P-P0} with $\beta=0$ and $\alpha=1.9$ (blue), $\alpha=2.4$ (yellow), and $\alpha=2.3$ (red). The symbols indicate the result from simulations in which both the windbreaks and wind turbines are included.}
\label{fig.linear-law}
\end{figure}

Figure~\ref{fig.linear-law}(a) shows the streamwise flow pattern for flow over a windbreak with a height $h/D=0.24$ and porosity $\eta=0.03$ (case 1 in Table~\ref{tab1}). In contrast to the case shown in Fig.~\ref{fig.validation}(c) we see the formation of a recirculation zone. The recirculation length, defined as the streamwise distance of the last semi-saddle point on the ground, is about $10 h$ \citep{fan97}. Figure~\ref{fig.linear-law}(a) also indicates the expected power increase based on the streamwise velocity profile over the windbreak using the first equality in Eq.~\er{eq.P-P0}. The figure shows that the maximum power production increase is predicted at $x_t/h = 4\sim 6$. To further illustrate these predictions, we show the expected power increase as a function of the windbreak height $h/z_h$ for different $x_t$ in Fig.~\ref{fig.linear-law}(b). This figure shows that the expected power production increase is highest for windbreaks of height $h/z_h \sim 0.25-0.35$, and depends on the distance between the windbreak and the wind turbine.

To assess the accuracy of these predictions, we compare them with the results from simulations in which both the wind turbine and windbreak are considered. In agreement with experimental results \citep{fan97, don07, tob17a} we find that for low windbreaks ($h/z_h \le 0.12$), the normalized power production for the first row agrees well with these predictions. However, we find that for high windbreaks, especially for large $x_t$, the power production increase obtained in the simulations with windbreaks does not necessarily agree with the simple estimates. The reason is that the turbine influences the flow development over the windbreak. The presence of such non-linear interactions emphasizes the need to perform detailed simulations, which can capture these effects. The error bars in the figure indicate the standard deviation of the power output, which shows that the power fluctuations increase with windbreak height.

\subsection{Effect of windbreaks on wind farm performance} \lb{sec.windfarm}

\begin{figure} [!tb] 
\centering
\begin{overpic}[width=0.263\textwidth]{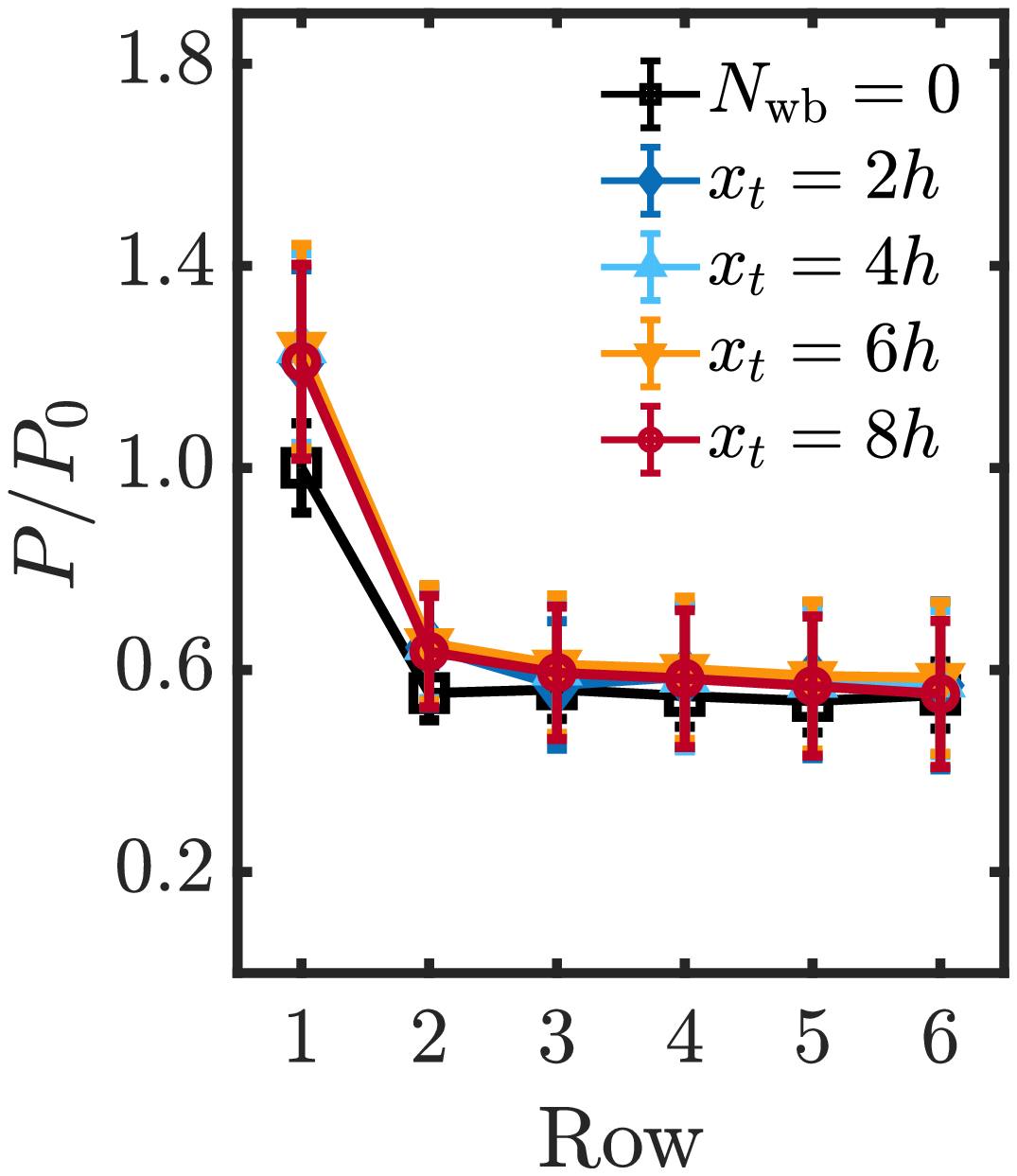} 
\put(0,91){$(a)$}
\end{overpic} 
\begin{overpic}[width=0.21\textwidth]{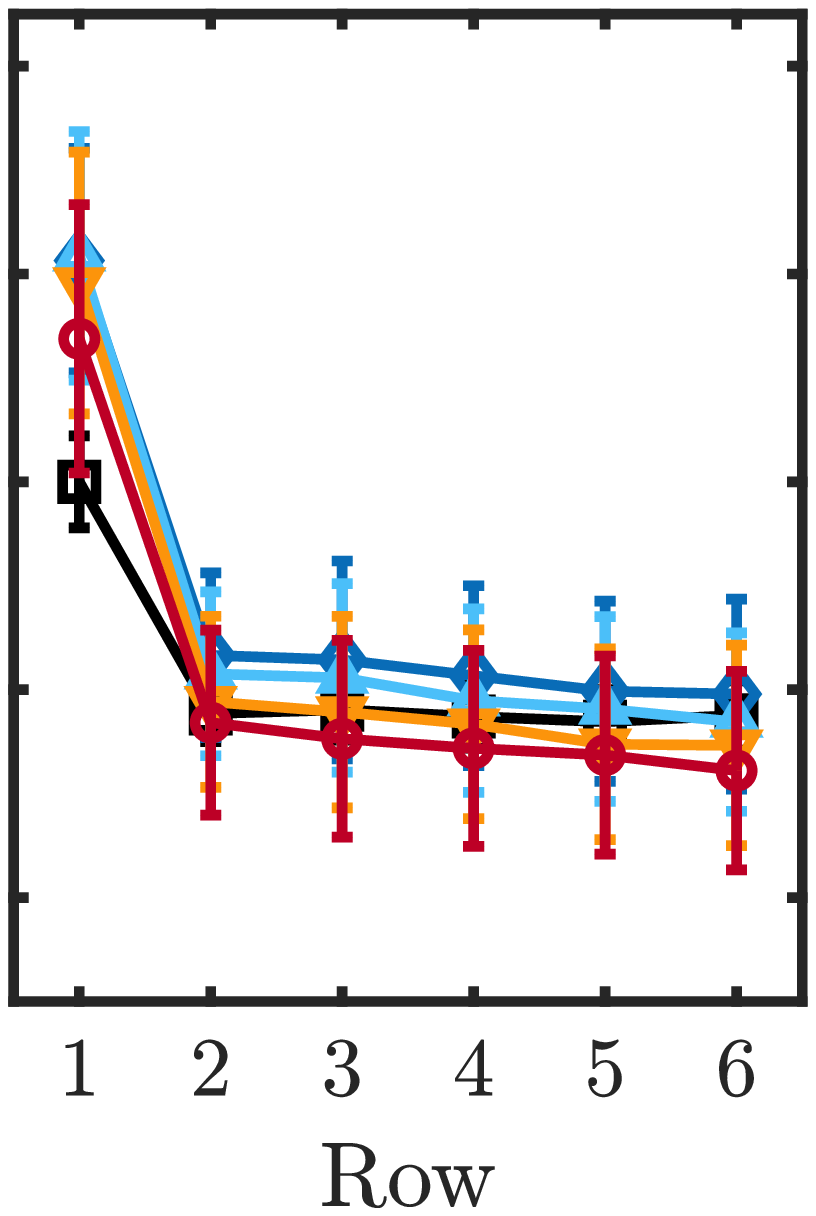} 
\put(-4,89){$(b)$}
\end{overpic}
\begin{overpic}[width=0.21\textwidth]{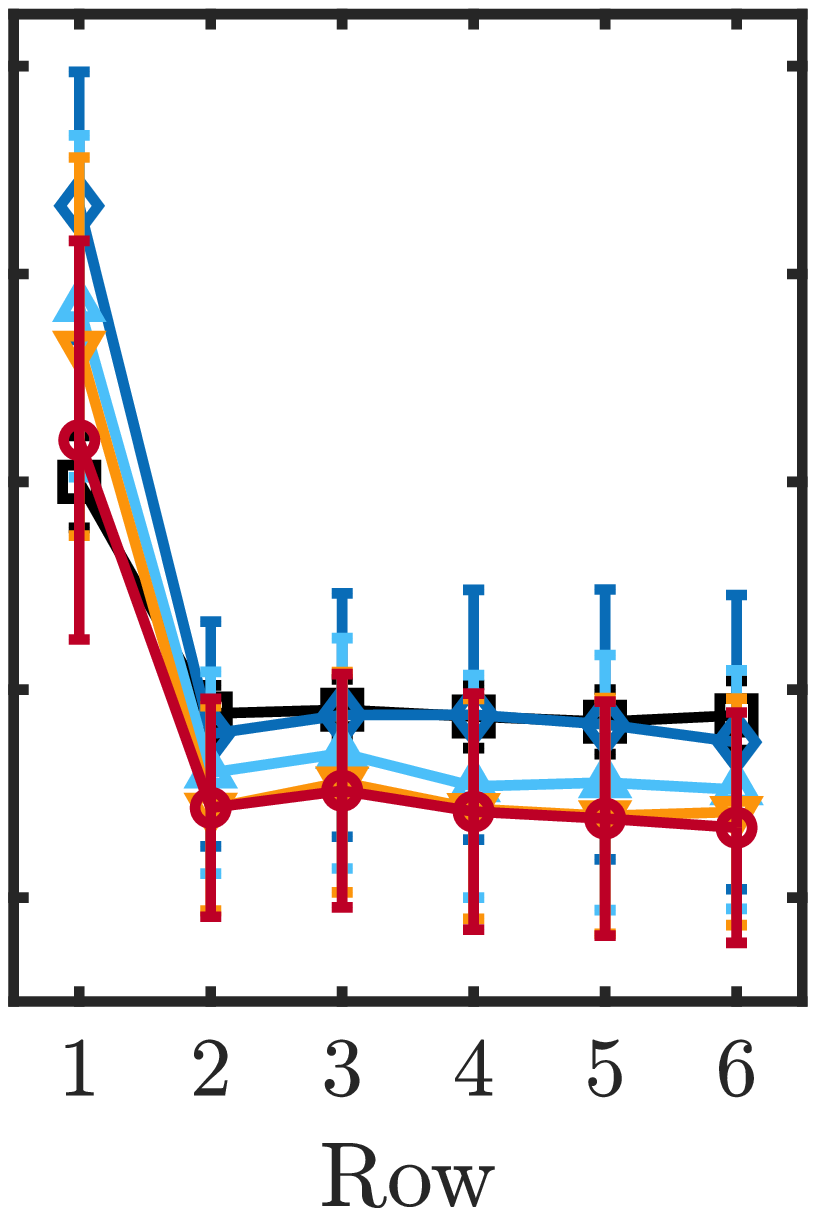} 
\put(-4,89){$(c)$}
\end{overpic}
\begin{overpic}[width=0.21\textwidth]{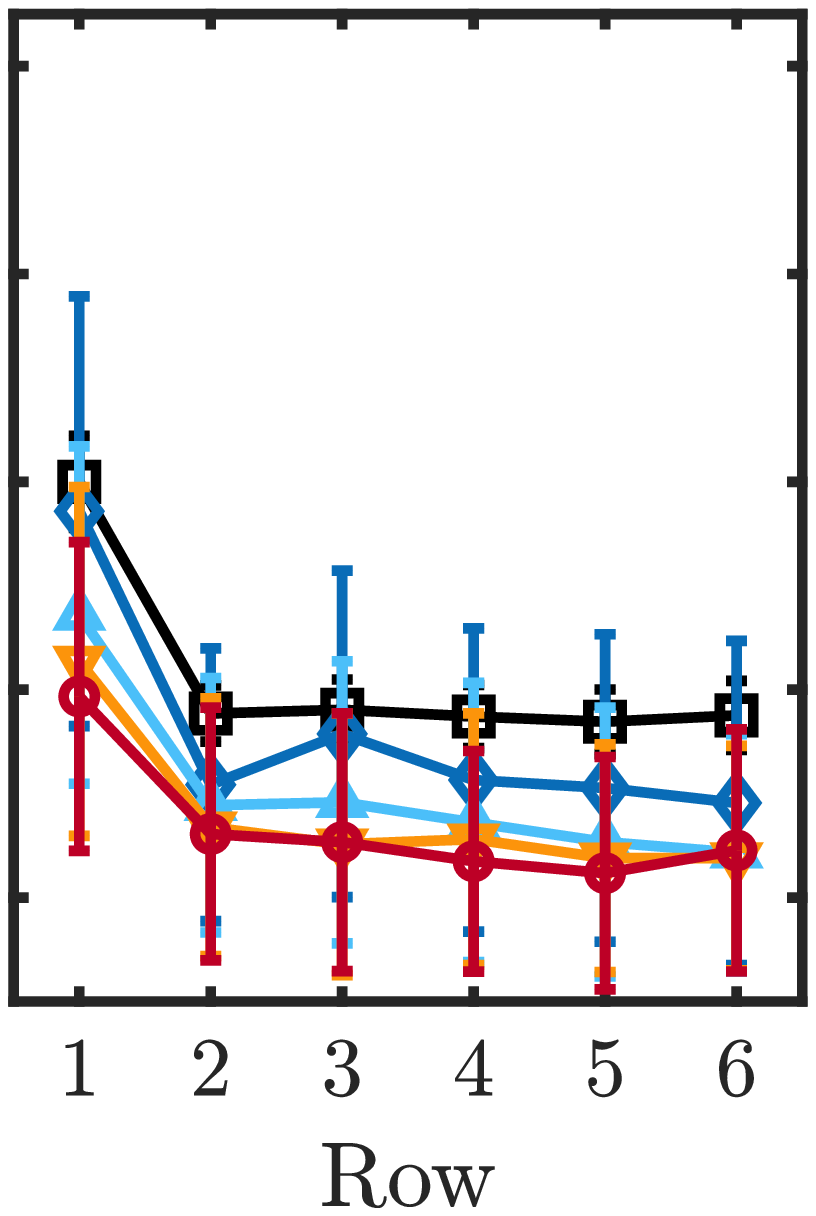} 
\put(-4,89){$(d)$}
\end{overpic}
\caption{The normalized power production $P/P_0$ as a function of downwind position for different $x_t$ (see figure \ref{fig.sketch}), where $P_{0}$ is the power production of a stand-alone wind turbine without windbreak. The windbreak porosity is $\eta=0.03$ and its height is (a) $h/z_h=0.12$, (b) $h/z_h=0.24$, (c) $h/z_h=0.36$, and (d) $h/z_h=0.48$. Values are also given in Table~\ref{tab1}.}
\label{fig.power-height}
\end{figure}

Figure~\ref{fig.power-height} shows the time-averaged normalized power production $P/P_{0}$ as a function of downwind position for different windbreak heights and locations, where $P_{0}$ is the power production of a stand-alone wind turbine without windbreak. Figure~\ref{fig.power-height}(a) shows that low windbreaks ($h/z_h=0.12$) increase the power production of each turbine in the wind farm compared to the corresponding reference case. This shows that windbreaks can be successfully employed to increase the power production of a wind farm. Figure \ref{fig.power-height}(b) shows that windbreaks of intermediate height ($h/z_h=0.24$) increase the power production of all rows when $x_t\le4h$. However, Fig.~\ref{fig.power-height}(b,c) also show that windbreaks of intermediate heights ($h/z_h=0.24$ and $h/z_h=0.36$) can significantly affect the performance of turbines further downwind. Figure~\ref{fig.power-height}(d) shows that the highest windbreaks ($h/z_h=0.48$) have a strong negative effect on the wind farm performance.

\begin{figure} [!tb]
\centering
\begin{overpic}[width=0.4\textwidth]{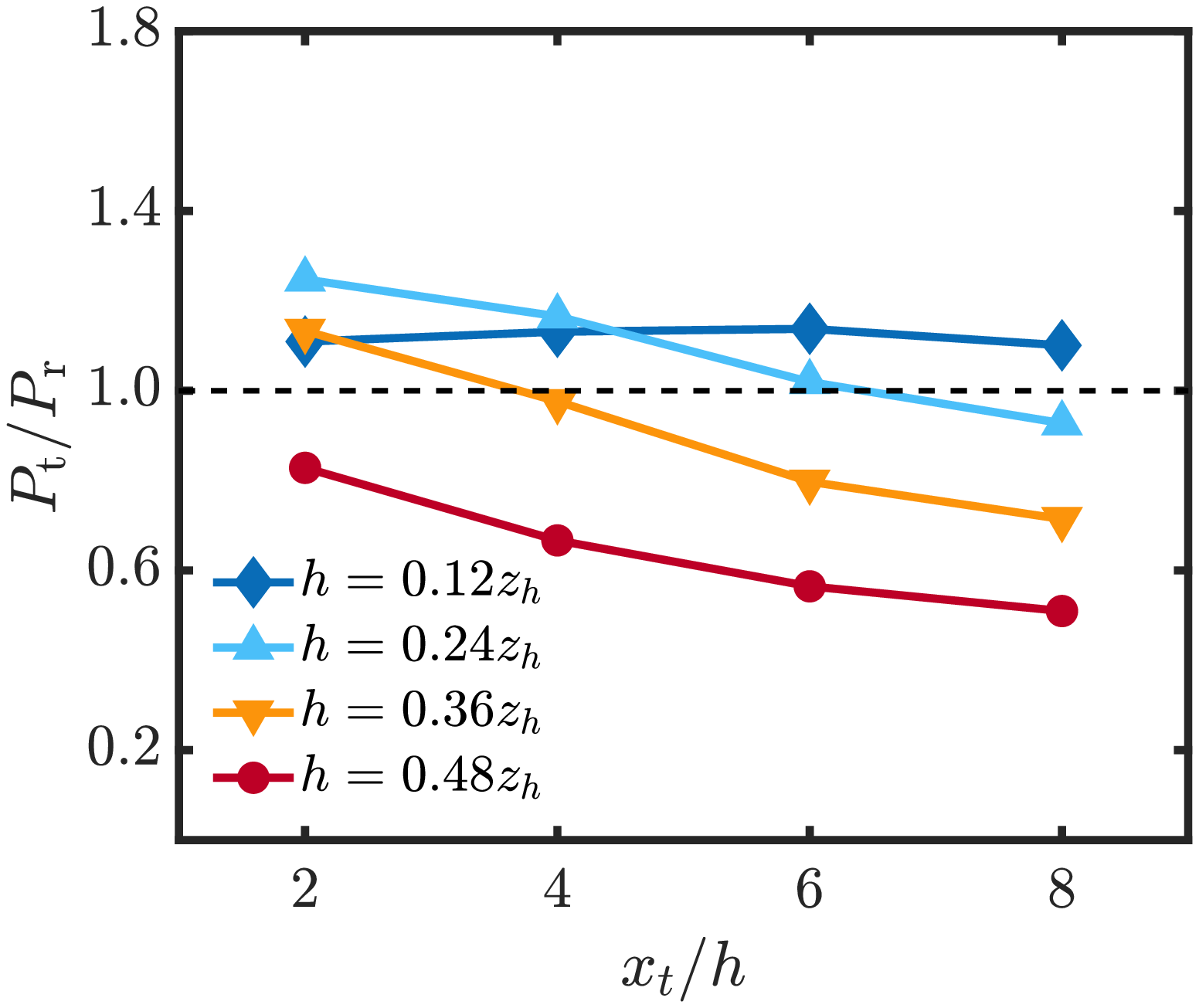} 
\put(2,73.5){$(a)$}
\end{overpic} 
\begin{overpic}[width=0.4\textwidth]{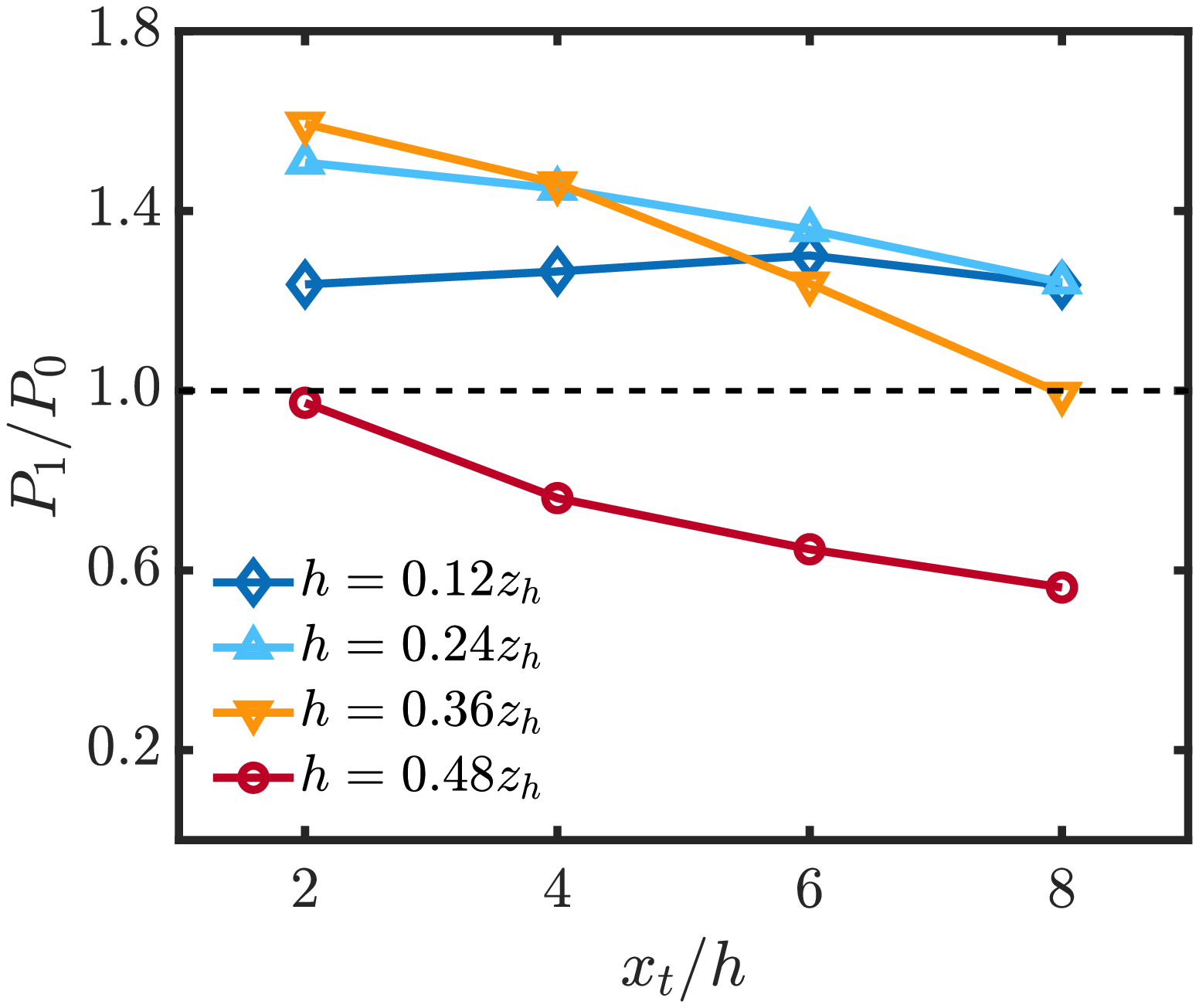} 
\put(2,73.5){$(b)$}
\end{overpic}
\caption{The normalized (a) wind farm $P_{\rm t}/P_{\rm r}$ and (b) first row $P_1/P_{0}$ power production as a function of the distance $x_t/h$ (see figure~\ref{fig.sketch}). Here $P_{\rm r}$ and $P_{0}$ are the power production of a wind farm and a stand-alone wind turbine without windbreak, respectively. The windbreak porosity is $\eta=0.03$. }
\label{fig.power-xt}
\end{figure}

To further analyze the effect of the windbreaks on the wind farm performance, we show the normalized time-averaged power production of the various wind farms with windbreaks in Fig.~\ref{fig.power-xt}(a). Figure~\ref{fig.power-xt}(b) shows the normalized performance of the first row. A comparison between both panels reveals that the front row benefits more from the windbreaks than the entire wind farm. For example, Fig.~\ref{fig.power-xt}(a) shows that low windbreaks ($h/z_h=0.12$) increase the wind farm production by $10\%$ to $14\%$, while the same windbreak increases the power production of the first row by $20\%$ to $25\%$. The spacing between the windbreak and wind turbine has a relatively small effect on the wind farm production for these low windbreaks. Windbreaks of intermediate height ($h/z_h=0.24$) can increase the wind farm power production more than low windbreaks. However, for these windbreaks, wind farm production strongly depends on the spacing between the windbreak and wind turbine. The wind farm production is usually reduced for high windbreaks, which is more pronounced for $h/z_h=0.48$ than for $h/z_h=0.36$. It is crucial to note that windbreaks that can increase the power production of the first row can reduce the power production of a wind farm.

\begin{figure} [!tb]
\centering
\begin{overpic}[width=0.3\textwidth]{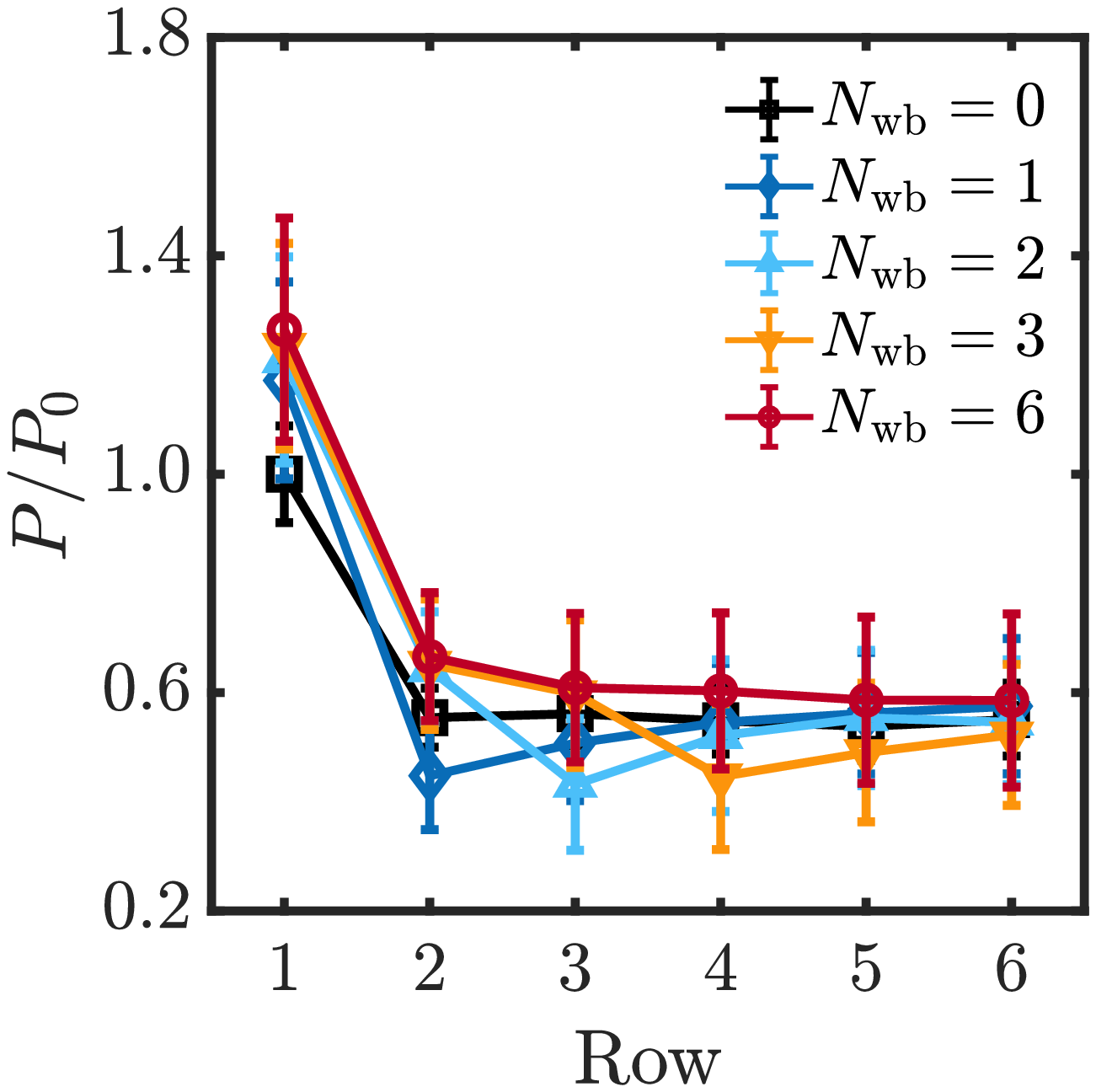} 
\put(0,88){$(a)$}
\end{overpic} 
\begin{overpic}[width=0.3\textwidth]{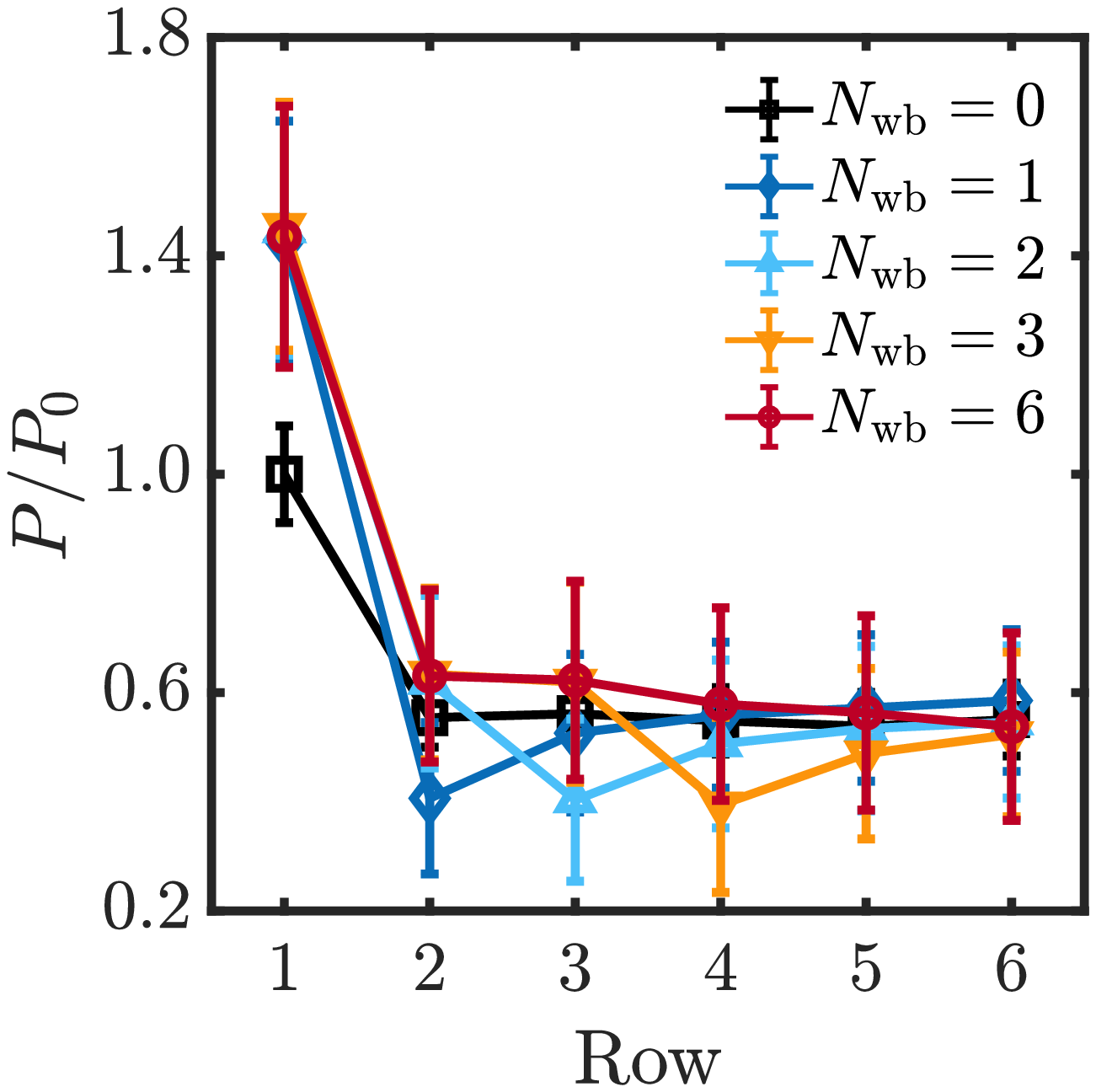} 
\put(0,88){$(b)$}
\end{overpic}
\begin{overpic}[width=0.3\textwidth]{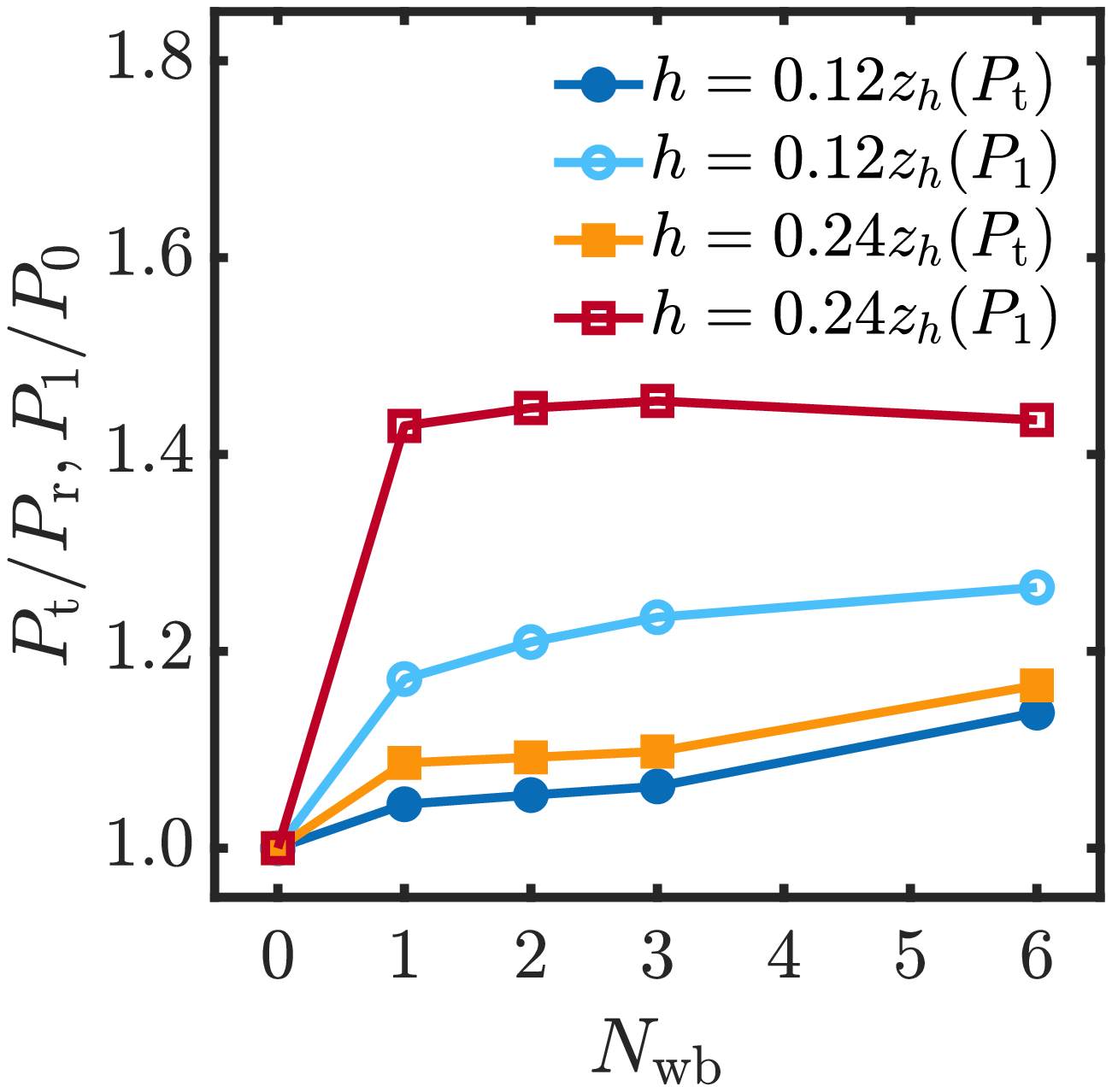} 
\put(0,88){$(c)$}
\end{overpic} 
\caption{(a,b) The normalized wind turbine power production $P/P_0$ as a function of downwind position using a different number of windbreaks $N_{\rm wb}$ with $\eta=0.03$ located at (a) $x_t/h=6, h/z_h=0.12$ and (b) $x_t/h=4, h/z_h=0.24$. (c) The corresponding normalized wind farm power production ($P_{\rm t}/P_{\rm r}$, filled symbols) and the first row power production ($P_1/P_0$, open symbols) as a function of the number of windbreaks. }
\label{fig.power-number}
\end{figure}

One may wonder how a windbreak that increases the performance of a single row may reduce the performance of a wind farm. To better understand this, we performed simulations with windbreaks installed upwind of only the first $N_{\rm wb}$ rows, where $N_{\rm wb}$ is the total number of windbreaks. In Fig.~\ref{fig.power-number} these cases are compared to the reference case without windbreaks ($N_{\rm wb}=0$) and the wind farm with windbreaks installed in front of every row ($N_{\rm wb}=6$). The figure shows that installing a windbreak in front of the first row ($N_{\rm wb}=1$) results in a substantial increase in the production of the first row. However, the wake created behind this windbreak significantly reduces the power production of the second row. Installing a windbreak in front of the second row ($N_{\rm wb}=2$) increases the power production of the second row compared to the $N_{\rm wb}=1$ case. However, due to the wake effect of the first windbreak, the second row's power production is only slightly higher than for the reference case without windbreaks. A similar effect is observed for the third row in the $N_{\rm wb}=3$ case. Figure~\ref{fig.power-number}(c) shows that the wind farm power output monotonically increases with the number of installed windbreaks. However, the front row benefits most from the windbreak due to the effect described above. This means that the effect of windbreaks on wind farm performance cannot be readily estimated. Note that $N_{\rm wb}$ has only a negligible effect on the first row production for high windbreaks $(h=0.24z_h)$. For low windbreaks $(h=0.12z_h)$ the the first row increases when windbreak are placed further downwind in the wind farm. 

\begin{figure}  [!tb]
\centering
\begin{overpic}[width=0.8\textwidth]{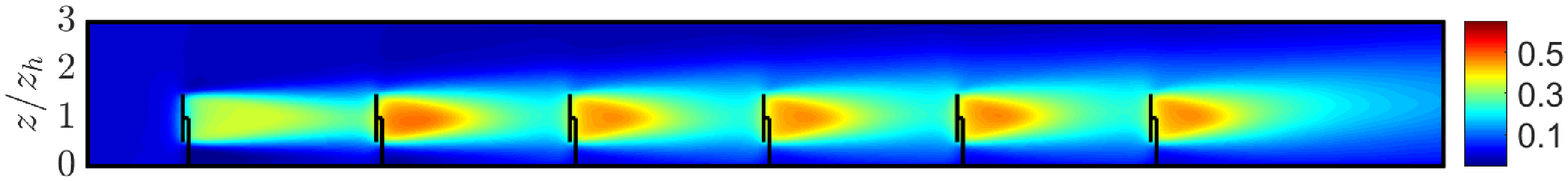} 
\put(0,10){$(a)$}
\end{overpic} 
\begin{overpic}[width=0.8\textwidth]{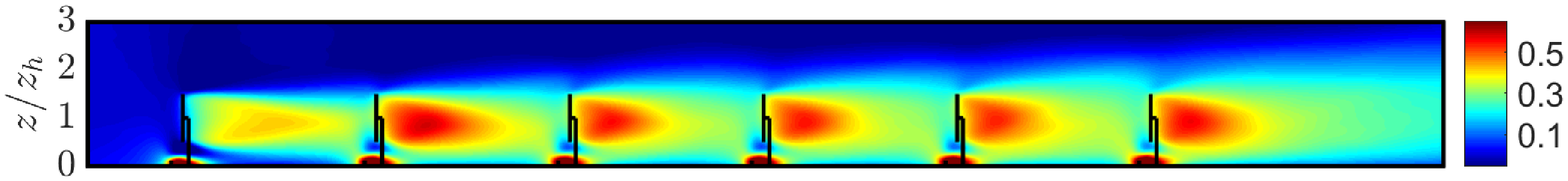} 
\put(0,10){$(b)$}
\end{overpic} 
\begin{overpic}[width=0.8\textwidth]{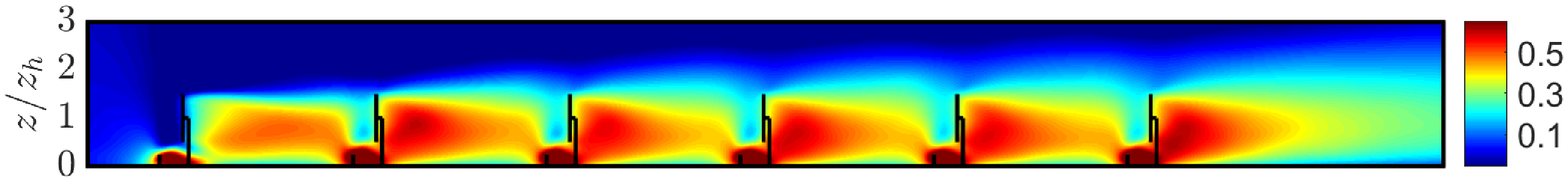} 
\put(0,10){$(c)$}
\end{overpic} 
\begin{overpic}[width=0.8\textwidth]{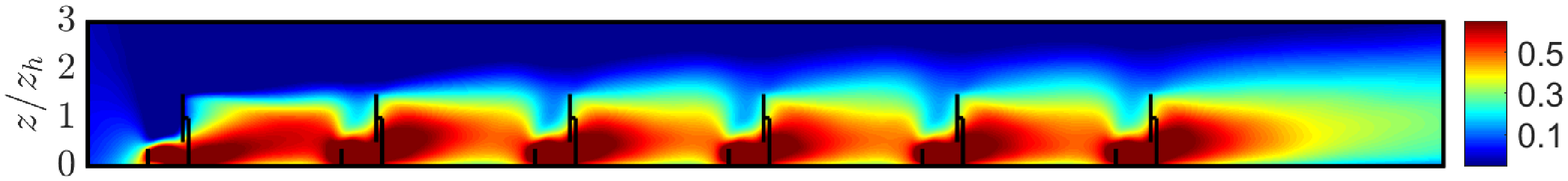} 
\put(0,10){$(d)$}
\end{overpic} 
\begin{overpic}[width=0.8\textwidth]{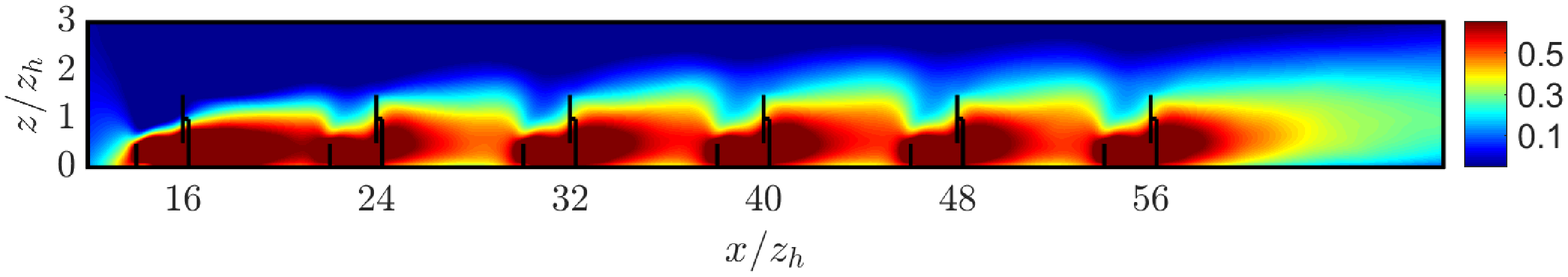} 
\put(0,15){$(e)$}
\end{overpic} 
\caption{Normalized streamwise velocity deficit $(U-U_{\rm ABL})/U_{\rm hub}$ in the mid-plane of the turbine. The distance between the windbreak and turbine is $x_t/h=4$. The windbreak porosity is $\eta=0.03$ and its height is (a) $h/z_h=0$, (b) $h/z_h=0.12$, (c) $h/z_h=0.24$, (d) $h/z_h=0.36$, and (e) $h/z_h=0.48$. The solid lines indicate the turbine and windbreak locations.}
\label{fig.udef}
\end{figure}

To analyze the effect of the windbreaks on the flow inside the wind farm in more detail, we show the time-averaged normalized streamwise velocity deficit $(U-U_{\rm ABL})/U_{\rm hub}$ in the vertical mid-plane of the turbine, where $U$ is the streamwise velocity in the wind farm, $U_{\rm ABL}$ is the incoming wind speed profile, and $U_{\rm hub}=U_{\rm ABL}(z_h)$ is the incoming wind speed at hub-height in Fig.~\ref{fig.udef}. For windbreaks with low and intermediate heights (Figs.~\ref{fig.udef}(b,c)), the turbine is not located in the windbreak wake region. Instead, the speed-up over the windbreak is the main effect that influences the power production of the first row. Figures~\ref{fig.udef}(a-c) show that the wind turbine wakes recover slower when the windbreaks are higher. The reason for this is the adverse pressure gradient that is created behind each windbreak (for the contribution of pressure on the power output, see Fig.~\ref{fig.budget}(d)). This phenomenon is very similar to the findings by \citet{sha18}, who showed that adverse pressure gradients induced by a hill negatively impact wake recovery.

\begin{figure}  [!tb]
\centering
\begin{overpic}[width=0.8\textwidth]{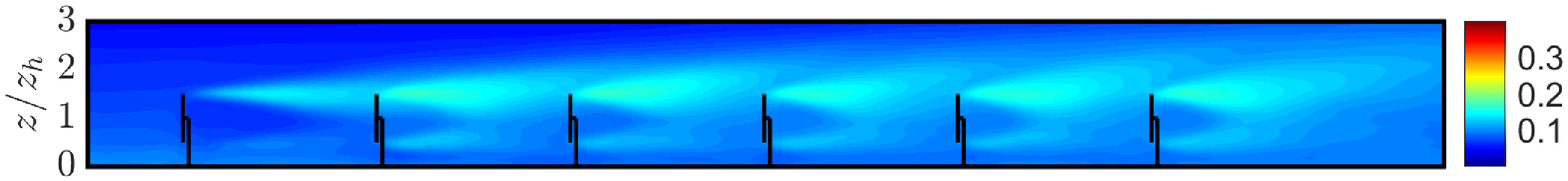} 
\put(0,10){$(a)$}
\end{overpic} 
\begin{overpic}[width=0.8\textwidth]{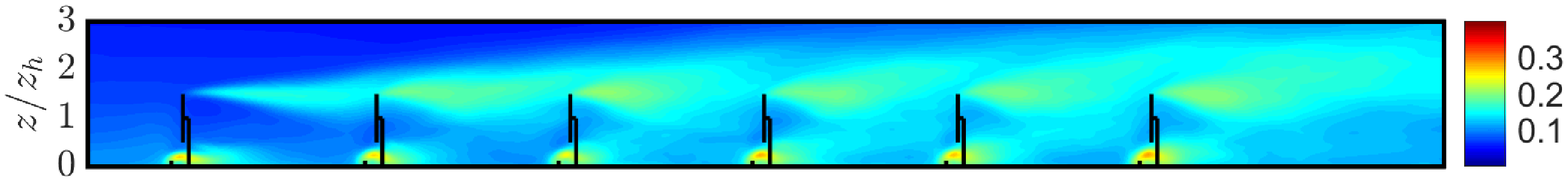} 
\put(0,10){$(b)$}
\end{overpic} 
\begin{overpic}[width=0.8\textwidth]{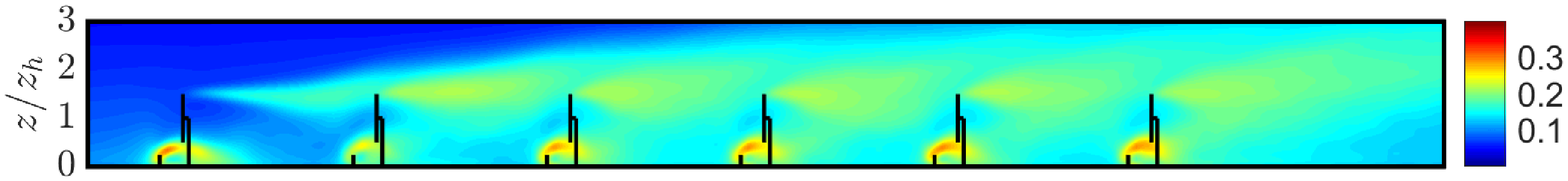} 
\put(0,10){$(c)$}
\end{overpic}
\begin{overpic}[width=0.8\textwidth]{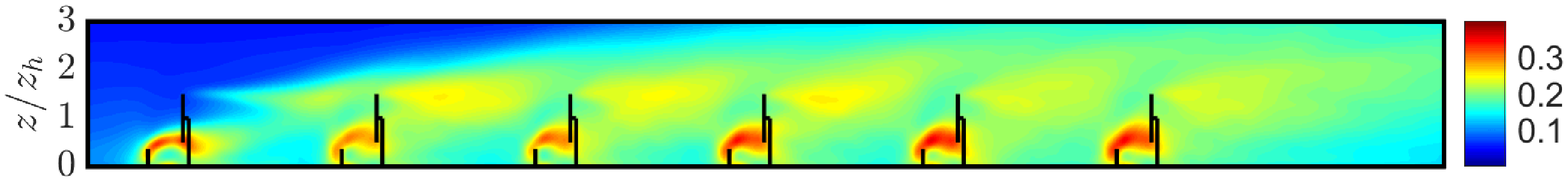} 
\put(0,10){$(d)$}
\end{overpic} 
\begin{overpic}[width=0.8\textwidth]{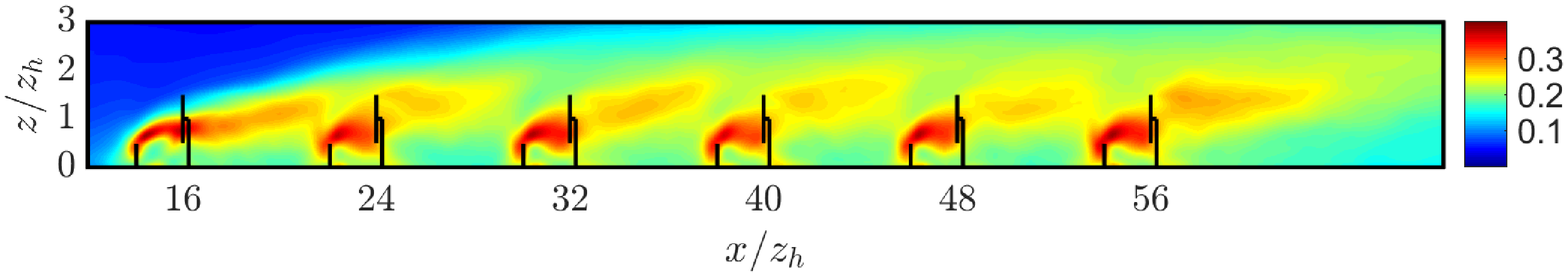} 
\put(0,15){$(e)$}
\end{overpic} 
\caption{Normalized streamwise velocity variance $\sigma_u/U_{\rm hub}$ in the mid-plane of the turbine. The distance between the windbreak and turbine is $x_t/h=4$. The windbreak porosity is $\eta=0.03$ and its height is (a) $h/z_h=0$, (b) $h/z_h=0.12$, (c) $h/z_h=0.24$, (d) $h/z_h=0.36$, and (e) $h/z_h=0.48$. The solid lines indicate the turbine and windbreak locations.}
\label{fig.urms}
\end{figure}

It is well known that wake recovery also depends on the turbulence intensity. In Fig.~\ref{fig.urms} we shows the normalized streamwise velocity variance $\sigma_u/U_{\rm hub}$. Figures \ref{fig.urms}(a-c) show that the increase in turbulence intensity at hub-height is limited. For high windbreaks with $h/z_h=0.36$ or $h/z_h=0.48$ the windbreak wake directly interacts with the wind turbine wake, see Figs.~\ref{fig.udef}(d,e) and Figs.~\ref{fig.urms}(d,e). Figures \ref{fig.urms}(d,e) show that these high windbreaks lead to a strong increase in the turbulence intensity, which reduces the turbine lifetime. As Fig.~\ref{fig.power-height} already showed that high windbreaks reduce the wind farm power production, it is clear that such high upwind obstacles should be avoided as much as possible.

\subsection{Kinetic energy budget analysis}\lb{sec.budget}

Wind farms extract kinetic energy from the ABL. At the first row of a wind farm, this energy is mainly extracted from the incoming wind by a horizontal flux of kinetic energy, while further downwind, it is entrained from the flow above the wind farm through the vertical energy transport associated with turbulence \citep{cal10,cal10b,ste14f,ste17,men19}. Here, we analyze the kinetic energy budget to clarify the underlying physical mechanism in wind farms with windbreaks.

Following \citet{abk14b}, we first perform time-averaging of the filtered momentum equation \citep{gad20, liu20, liu20d}, 
\beq\lb{eq.momentum2}
\pat_t \overline{ \widetilde{\pu} } + \overline{ \widetilde{\pu} } \cdot \na \overline{ \widetilde{\pu} } = \overline \pf_{\rm wt} + \overline \pf_{\rm wb} + \overline \pf_{p} - \na \overline{ \widetilde {p}} - \na \cdot \overline \pt - \na \cdot (\overline{ \widetilde{\pu}' \widetilde{\pu}' }),
\eeq
where the overline denotes time-averaging and $\overline{ \widetilde{\pu}' \widetilde{\pu}' } = \overline{ \widetilde{\pu} \widetilde{\pu} } - \overline{ \widetilde \pu } \, \overline{ \widetilde \pu }$ is the dispersive stress. Then, multiplying $\overline{ \widetilde{\pu} }$ on both sides of Eq.\ \er{eq.momentum2} and after some algebra, one obtains 
\beq\lb{eq.momentum4}
\pat_t k + \na \cdot \left( k \overline{ \widetilde{\pu} } + \overline{ \widetilde {p}} \, \overline{ \widetilde{\pu} } + \overline \pt \cdot\overline{ \widetilde{\pu} } + \overline{ \widetilde{\pu}' \widetilde{\pu}' } \cdot\overline{ \widetilde{\pu} } \right) = \left( \overline \pf_{\rm wt} + \overline \pf_{\rm wb} + \overline \pf_{p} \right) \cdot\overline{ \widetilde{\pu} } - \varepsilon_{\rm sgs} - \varepsilon_t,
\eeq 
where
\beq
k \equiv \frac{1}{2} \overline{ \widetilde{\pu} } \cdot \overline{ \widetilde{\pu} }, \quad 
\varepsilon_{\rm sgs} \equiv - \overline \pt : \na \overline{ \widetilde{\pu} }, \quad 
\varepsilon_t \equiv - \overline{ \widetilde{\pu}' \widetilde{\pu}' } : \na \overline{ \widetilde{\pu} }.
\eeq 
We take the integral around each turbine $V_T=[x_h-S_x/2, x_h+S_x/2] \times [y_h-S_y/2, y_h+S_y/2] \times [z_h-D/2, z_h+D/2]$ where $\px_h=(x_h,y_h,z_h)$ is the center location of turbine. We focus on the statistical stationary state in which
\beq\lb{eq.budget}
P_{\rm wt} = P_{\infty} + T_k + T_p + T_{\rm sgs} + T_t - D_{\rm sgs} - D_t.
\eeq
Here, $P_{\rm wt}$ and $P_\infty$ are the power generated by the turbine and pressure gradient, 
\beq
P_{\rm wt} = - \int_{V_T} \overline \pf_{\rm wt} \cdot\overline{ \widetilde{\pu} } \textrm{d} V, \quad
P_{\infty} = \int_{V_T} \overline \pf_{p} \cdot \overline{ \widetilde{\pu} } \textrm{d} V, \quad 
\eeq
$D_{\rm sgs}$ and $D_t$ are the total dissipation of turbulence stress and dispersive stress, 
\beq
D_{\rm sgs} = \int_{V_T} \overline \varepsilon_{\rm sgs} \textrm{d} V, \quad 
D_t = \int_{V_T} \overline \varepsilon_t \textrm{d} V,
\eeq 
$T_k$ and $T_p$ are the net fluxes of kinetic energy and modified pressure, 
\beq
T_{k} = -\int_{\pat V_T} k \overline{ \widetilde{\pu} } \cdot \pn \textrm{d} S, \quad 
T_{p} = -\int_{\pat V_T} \overline {\widetilde p} \, \overline{ \widetilde{\pu} } \cdot \pn \textrm{d} S
\eeq 
and $T_{\rm sgs}$ and $T_t$ are the net fluxes of turbulence stress and dispersive stress,
\beq
T_{\rm sgs} = -\int_{\pat V_T} \pn \cdot \overline \pt \cdot\overline{ \widetilde{\pu} } \textrm{d} S, \quad 
T_{t} = -\int_{\pat V_T} \pn \cdot \overline{ \widetilde{\pu}' \widetilde{\pu}' } \cdot\overline{ \widetilde{\pu} } \textrm{d} S,
\eeq 
with $\pat V_T$ denoting the boundary of the integral volume $V_T$. For infinite wind farms, the dominant term is $T_t$, which contributes to the integral of the vertical kinetic energy flux $-\overline{ \widetilde{u}' \widetilde{w}' } \, \overline{ \widetilde{u} }$ \citep{cal10,cal10b,ste14f,ste17}. However, as we will show below, the flow dynamics in wind farms with windbreaks are different.

\begin{figure}  [!tb]
\centering
\begin{overpic}[width=0.3\textwidth]{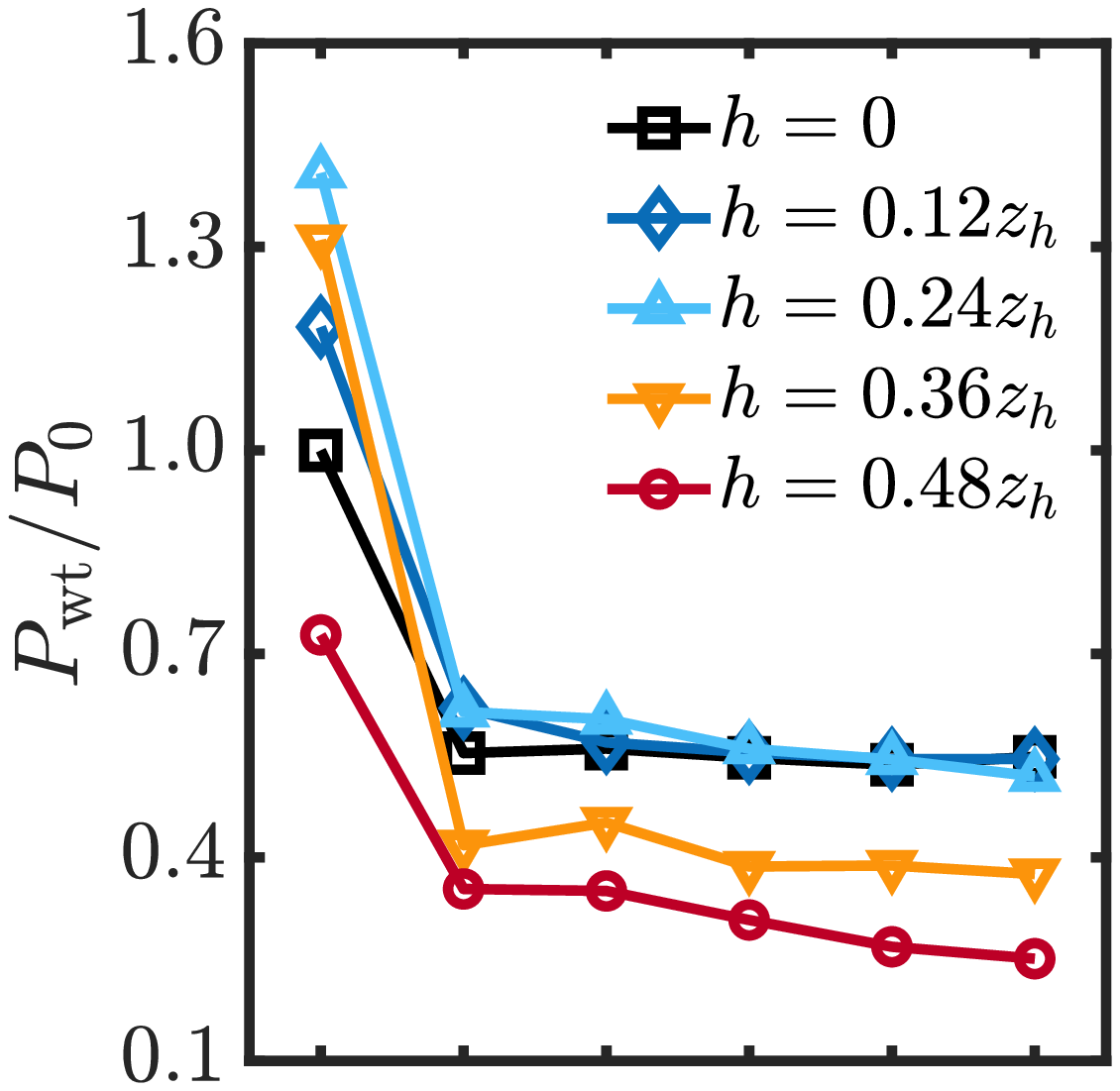} 
\put(0,83){$(a)$}
\end{overpic} 
\begin{overpic}[width=0.3\textwidth]{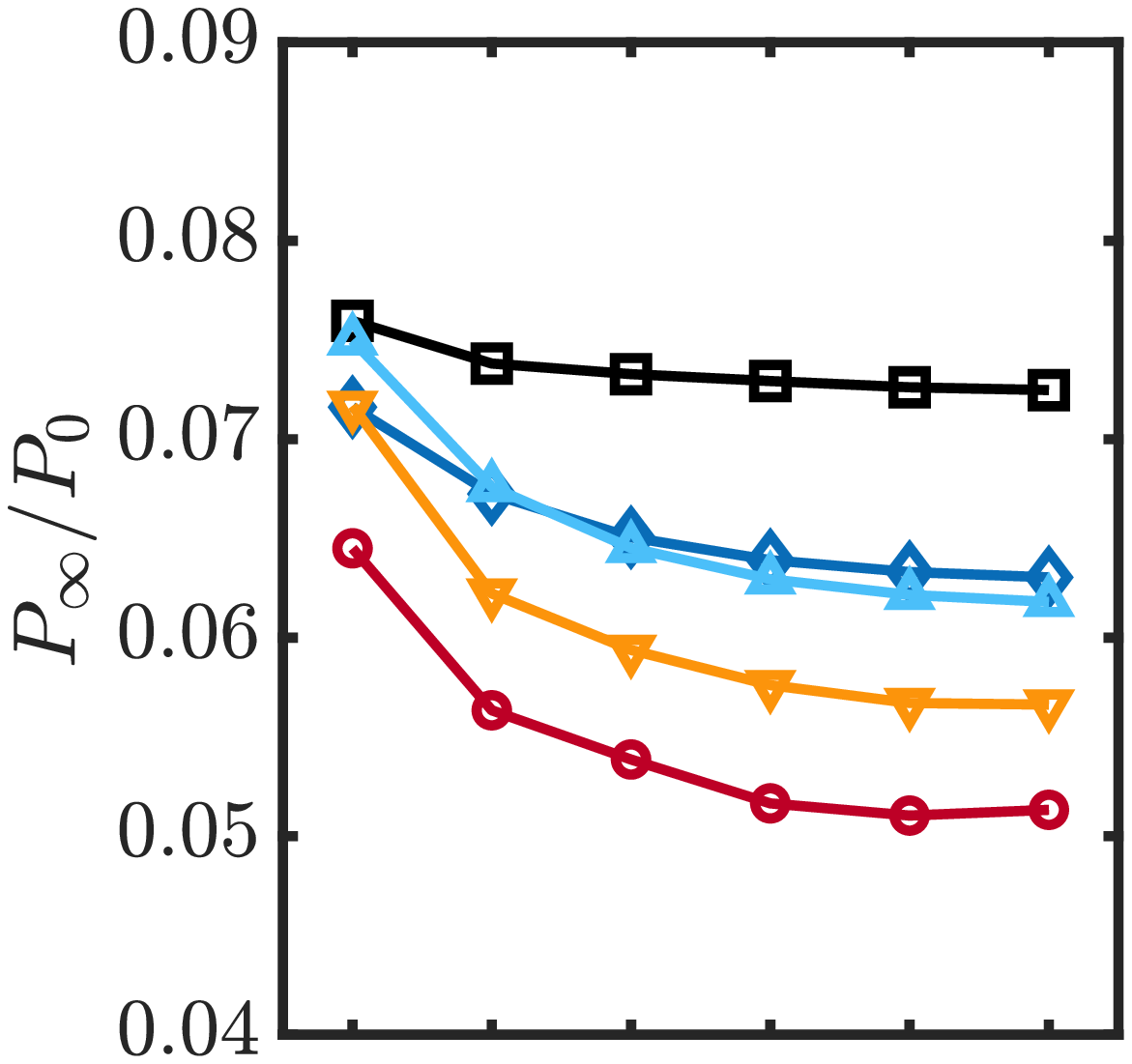} 
\put(0,83){$(b)$}
\end{overpic}
\begin{overpic}[width=0.3\textwidth]{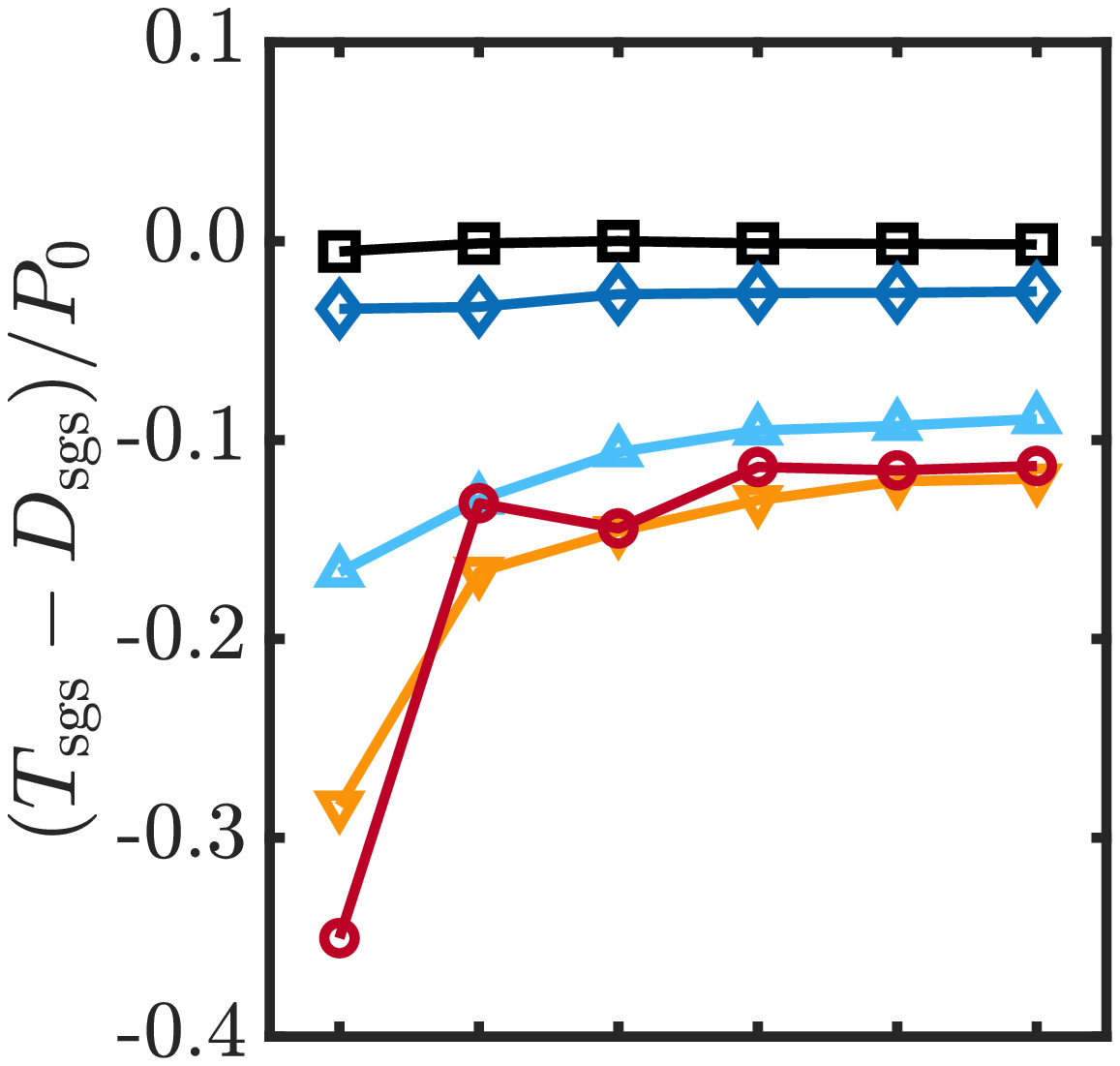} 
\put(0,83){$(c)$}
\end{overpic} 
\begin{overpic}[width=0.3\textwidth]{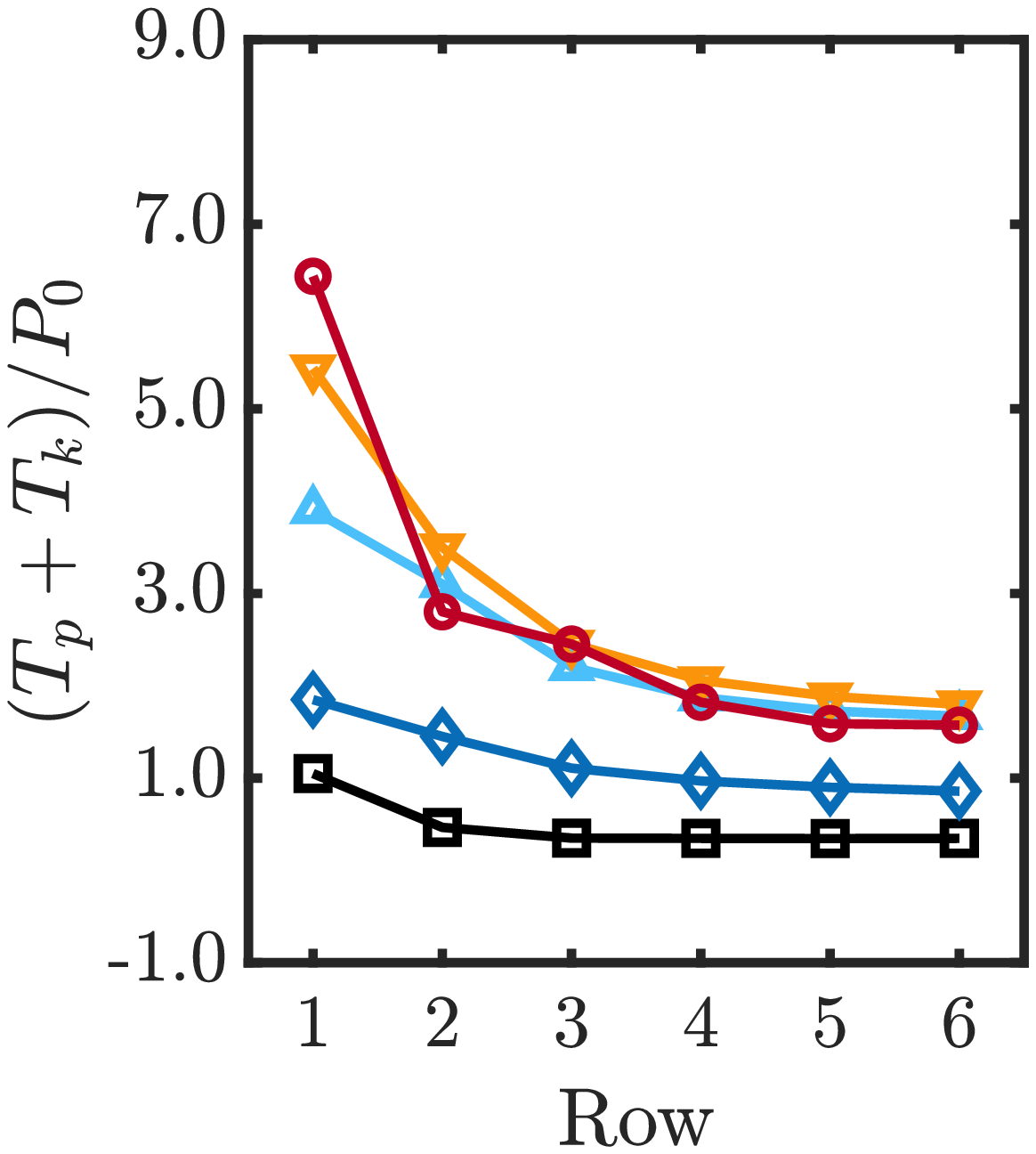} 
\put(0,92){$(d)$}
\end{overpic} 
\begin{overpic}[width=0.3\textwidth]{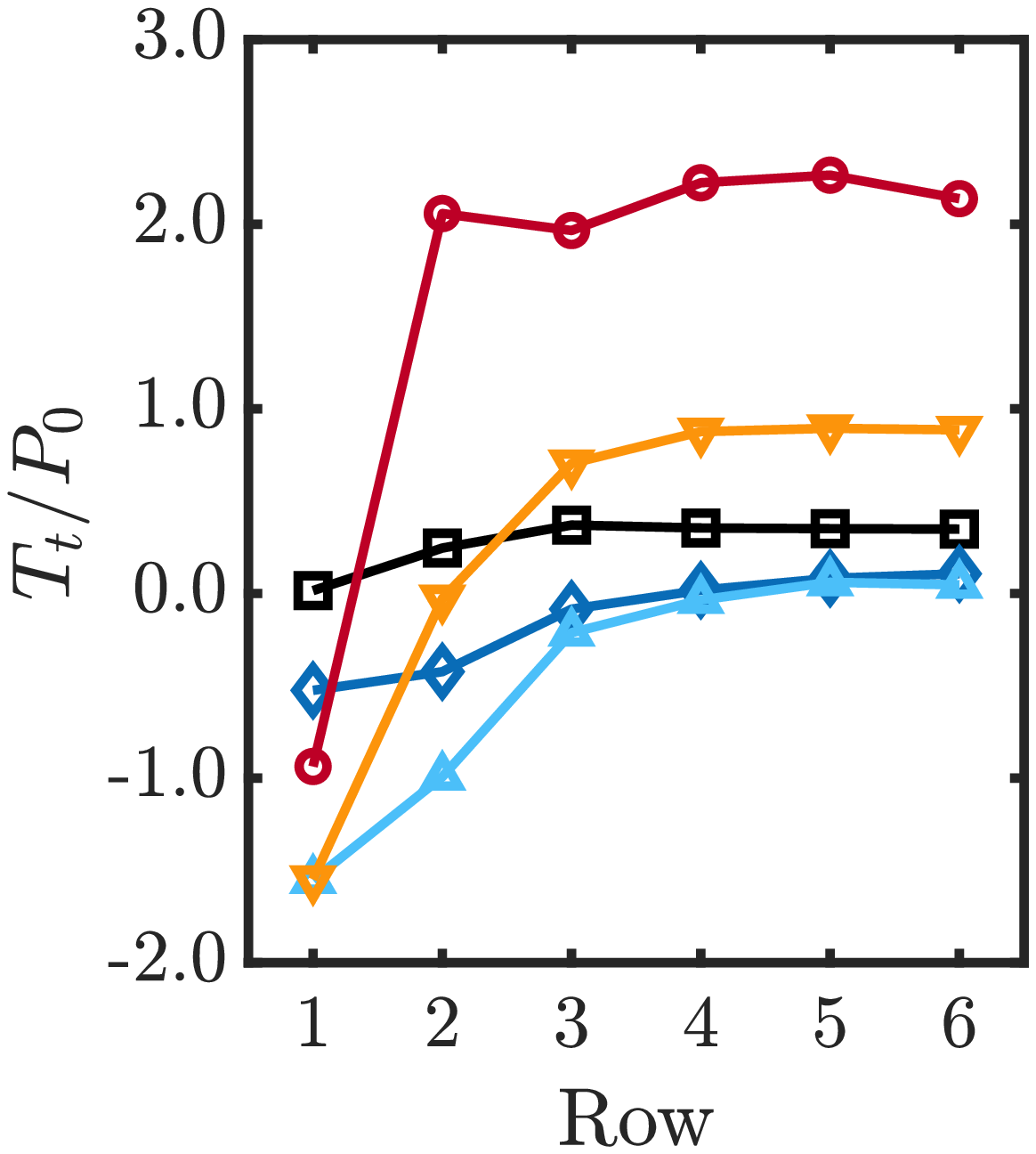} 
\put(0,92){$(e)$}
\end{overpic} 
\begin{overpic}[width=0.3\textwidth]{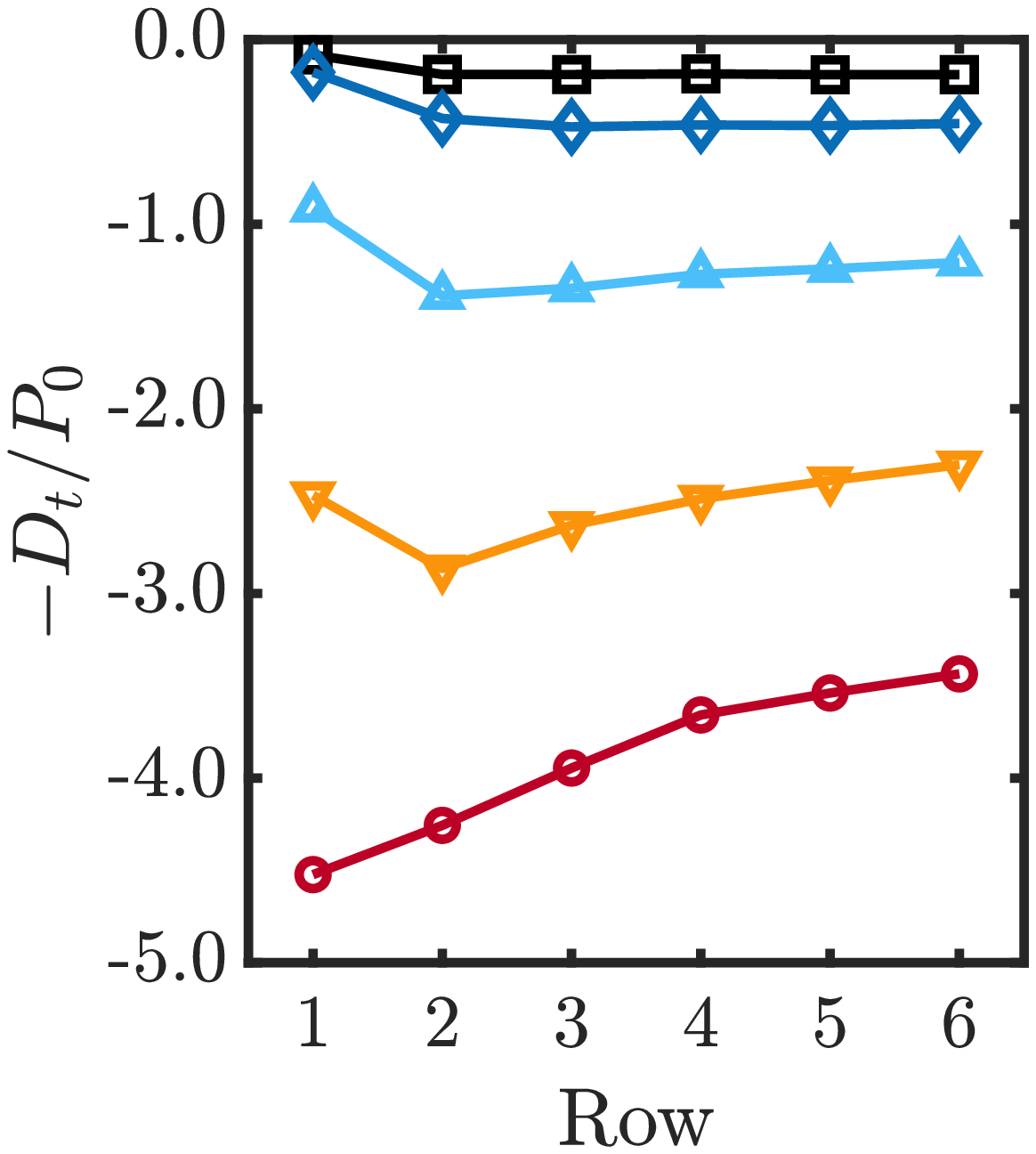} 
\put(0,92){$(f)$}
\end{overpic} 
\caption{Energy budget for wind farms with different windbreak height $h$ as a function of downwind position. The windbreak is located $x_t=4h$ upwind of the turbines. Each term in the energy budget equation \er{eq.budget} is normalized with the power production of an isolated wind turbine $P_0$. The windbreak is two-dimensional and has a porosity $\eta=0.03$.}
\label{fig.budget}
\end{figure}

Figure~\ref{fig.budget} shows the relative contribution of each term in the energy budget equation \er{eq.budget} as a function of downwind position for four different windbreak heights $h$ and a reference wind farm without windbreaks. We note that the data in Fig.~\ref{fig.budget}(a) are equivalent to the performance data shown in Fig.~\ref{fig.power-height}. As indicated by equation \er{eq.budget}, the power production of the wind turbine can be categorized into seven different sources and sinks. The contribution of the mean pressure gradient (Fig.~\ref{fig.budget}(b)) is always positive, and that of the dissipation term is always negative (Fig.~\ref{fig.budget}(f)). In contrast, the contribution of the transport term can be either positive or negative; see Fig.~\ref{fig.budget}(e). Figures~\ref{fig.budget}(b,c) show that the production term due to the pressure gradient, $P_\infty$, and the contribution of the turbulence stress, $T_{\rm sgs}-D_{\rm sgs}$, are usually one order of magnitude smaller than the leading order terms. Therefore, these terms are neglected in the following discussion. 

Figure~\ref{fig.budget}(d) shows that the contribution of the total pressure ($T_p+T_k$) is always positive and the dominant term at the leading edge of the wind farm. However, its magnitude decreases further downwind in the wind farm. A detailed analysis reveals that the horizontal total pressure flux, i.e.\ $-(\overline {\widetilde p}+k) \overline {\widetilde u}$, which represents the flow energy that can be extracted from the upwind wind, is the dominant contribution to this term. The figure shows that this term increases with windbreak height in the entrance region of the wind farm. However, it saturates with increasing windbreak height ($h/z_h>0.24$) further downwind. This is the first reason higher windbreaks are not effective in improving wind farm performance. 

Figure~\ref{fig.budget}(e) shows the contribution of the transport term of the dispersive stress $T_{t}$, which represents the amount of kinetic energy that is entrained into (positive) or out of (negative) the volume $V_T$. For the wind farm without windbreaks, this term is positive, and it increases further downwind in the wind farm. Since the dissipation term $D_t$ is always negative (see Fig.~\ref{fig.budget}(f)), the transport term $T_t$ provides a crucial contribution to the performance of large wind farms. This shows that for wind farms without windbreaks, the vertical kinetic energy flux ($-\overline{ \widetilde{u}' \widetilde{w}' } \, \overline{ \widetilde{u} }$) can be regarded as a measure of the amount of energy that can be entrained from outside the control volume $V_T$ in the fully developed wind farm regime.

\begin{figure} [!tb] 
\centering
\begin{overpic}[width=0.8\textwidth]{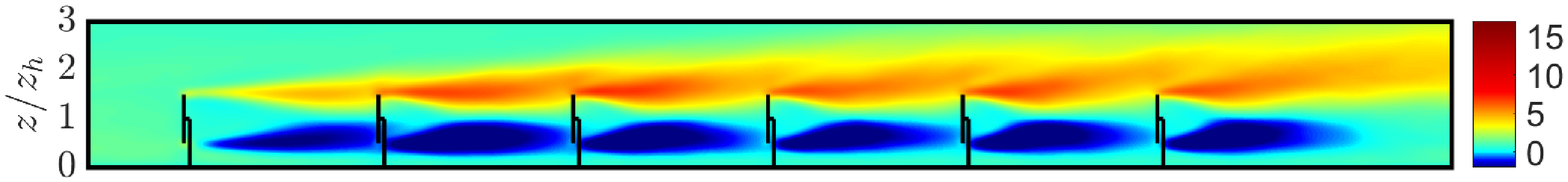} 
\put(0,10){$(a)$}
\end{overpic} 
\begin{overpic}[width=0.8\textwidth]{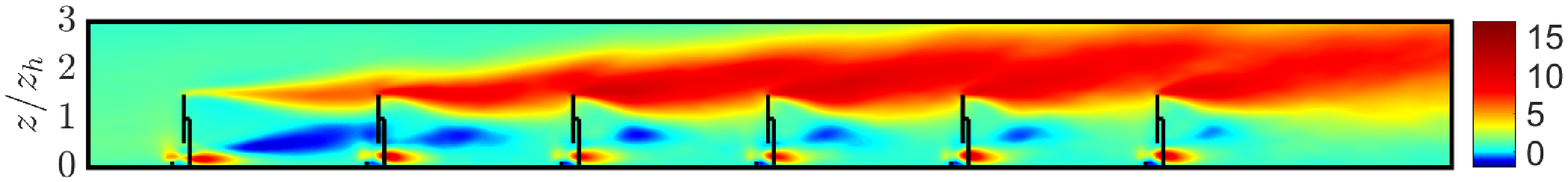} 
\put(0,10){$(b)$}
\end{overpic} 
\begin{overpic}[width=0.8\textwidth]{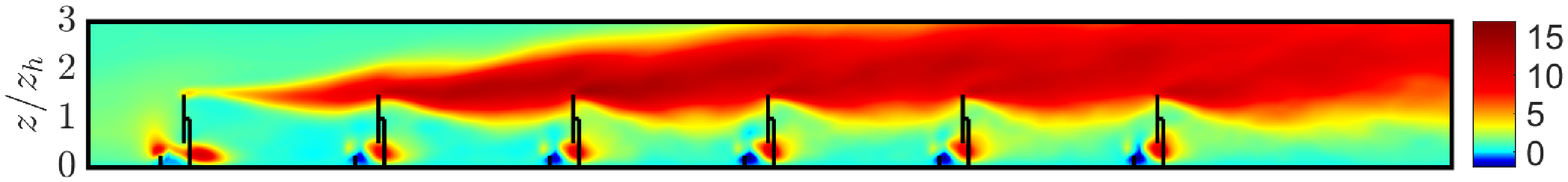} 
\put(0,10){$(c)$}
\end{overpic} 
\begin{overpic}[width=0.8\textwidth]{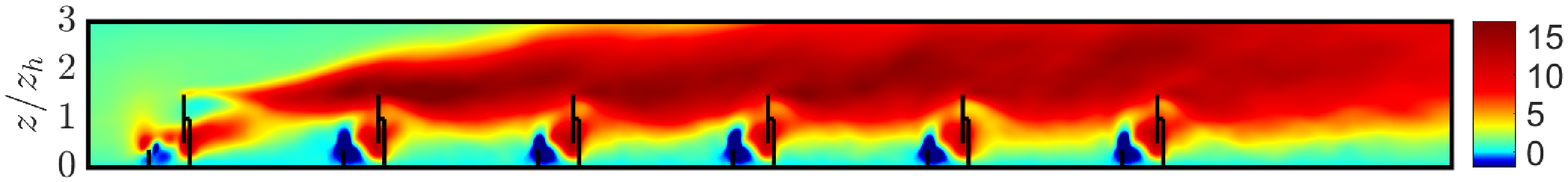} 
\put(0,10){$(d)$}
\end{overpic} 
\begin{overpic}[width=0.8\textwidth]{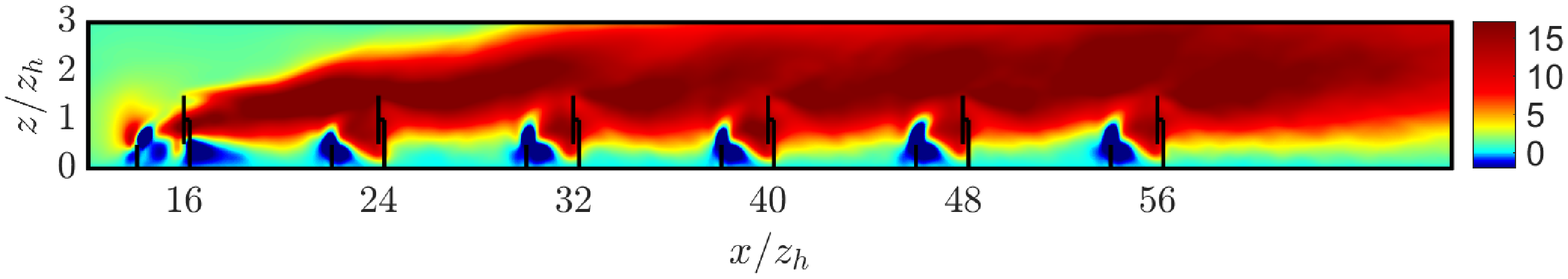} 
\put(0,15){$(e)$}
\end{overpic} 
\caption{Normalized vertical kinetic energy flux $-10^3 \overline{ \widetilde{u}' \widetilde{w}' } \, \overline{ \widetilde{u} } / U_{\rm hub}^3$ in the mid-plane of the turbine. The distance between the windbreak and turbine is $x_t/h=4$. The windbreak height is (a) $h/z_h=0$, (b) $h/z_h=0.12$, (c) $h/z_h=0.24$, (d) $h/z_h=0.36$, and (e) $h/z_h=0.48$. The windbreak porosity is $\eta=0.03$. The solid lines indicate the turbine and windbreak positions.}
\label{fig.flux}
\end{figure}

However, this scenario is not applicable for wind farms with windbreaks. To illustrate this, Fig.~\ref{fig.flux} shows the normalized vertical kinetic energy flux in the mid-plane of the turbine. For the reference wind farm without windbreaks, the flux is positive at $z=z_h+D/2$ and negative at $z=z_h-D/2$. This indicates that the wind turbine can extract energy from above or below. The flux above the turbine increases with the windbreak height. However, the flux at $z=z_h-D/2$ also increases significantly, implying that less kinetic energy can be entrained from below the turbine. For low and intermediate windbreaks, the flux increases more at $z=z_h-D/2$ than at $z=z_h+D/2$. As a result, the contribution of the vertical kinetic energy flux becomes negative, see Fig.~\ref{fig.budget}(e), due to which it hinders the performance of the turbines. For high windbreaks, the same happens for the first row. However, further downwind, the flux at $z=z_h+D/2$ becomes larger than at $z=z_h-D/2$ due to the significantly increased turbulence fluctuations, see Fig.~\ref{fig.urms}(d,e). Consequently, for higher windbreaks $(h/z_h\ge0.36)$ the contribution of the vertical kinetic energy flux becomes positive again further downwind, see Fig.~\ref{fig.budget}(e).

Figure~\ref{fig.budget}(f) shows that the contribution of the dissipation of the dispersive stress $D_{t}$ is always negative and becomes the dominant term with increasing windbreak height. This is consistent with the increase in the turbulence fluctuations shown in Fig.~\ref{fig.urms}. For high windbreaks, the increase in the dissipation term overwhelms the transport term $T_t$ and pressure terms $T_p+T_k$. This explains why high windbreaks reduce the power production of wind farms. 

Crucially, the above observation shows that the kinetic energy budget in wind farms is significantly affected by windbreaks. While it is commonly accepted that there is a significant correlation between the vertical kinetic energy $-\overline{ \widetilde{u}' \widetilde{w}' } \, \overline{ \widetilde{u} }$ and the power extraction by the turbines in large wind farms without windbreak \citep{cal10,cal10b,ste14f,ste17}, our results show that this completely changes when windbreaks are added.

\subsection{Effect of windbreak width and porosity}\lb{sec.porosity}

\begin{figure} 
\centering
\begin{overpic}[width=0.3\textwidth]{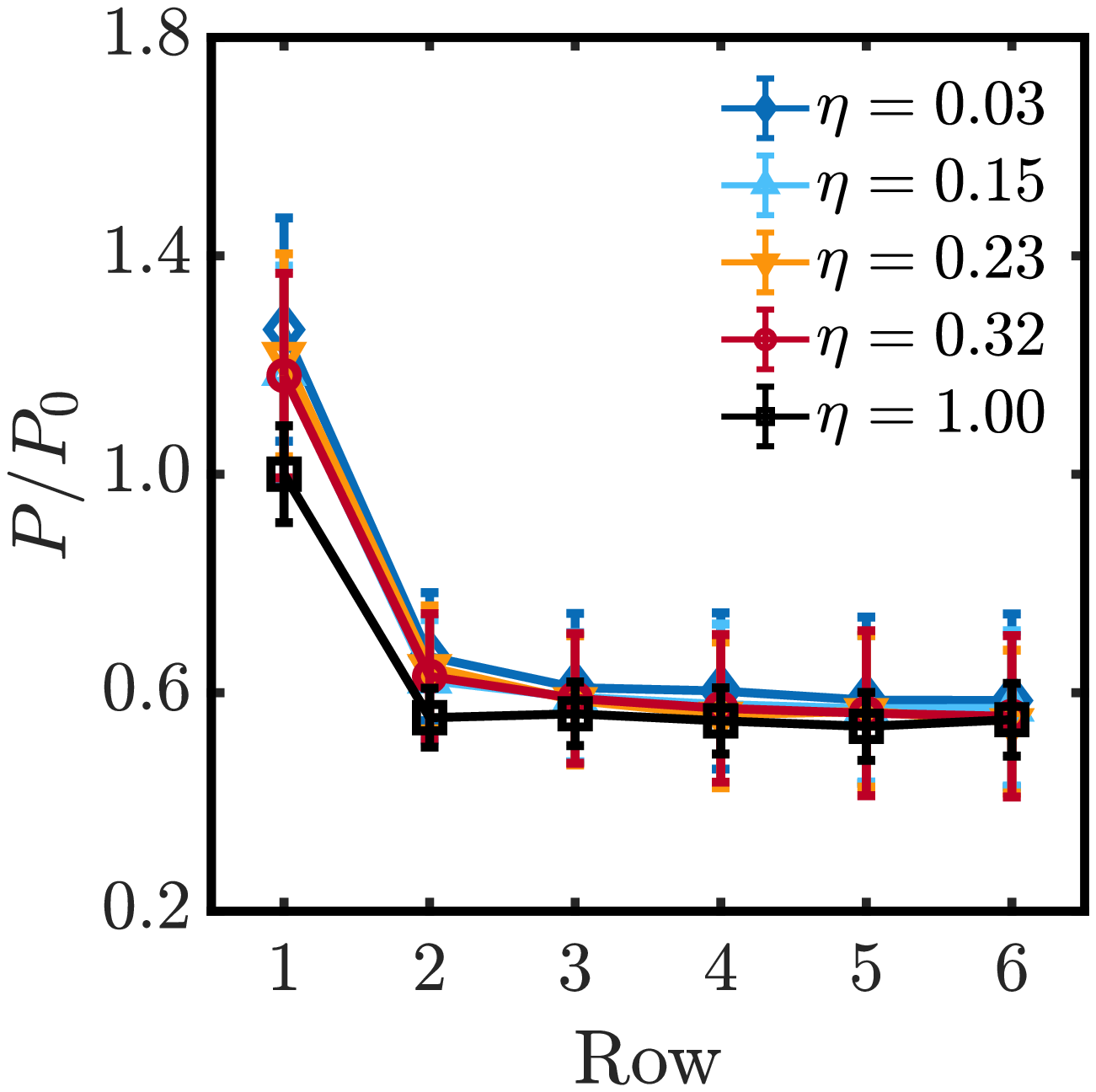} 
\put(0,88){$(a)$}
\end{overpic} 
\begin{overpic}[width=0.3\textwidth]{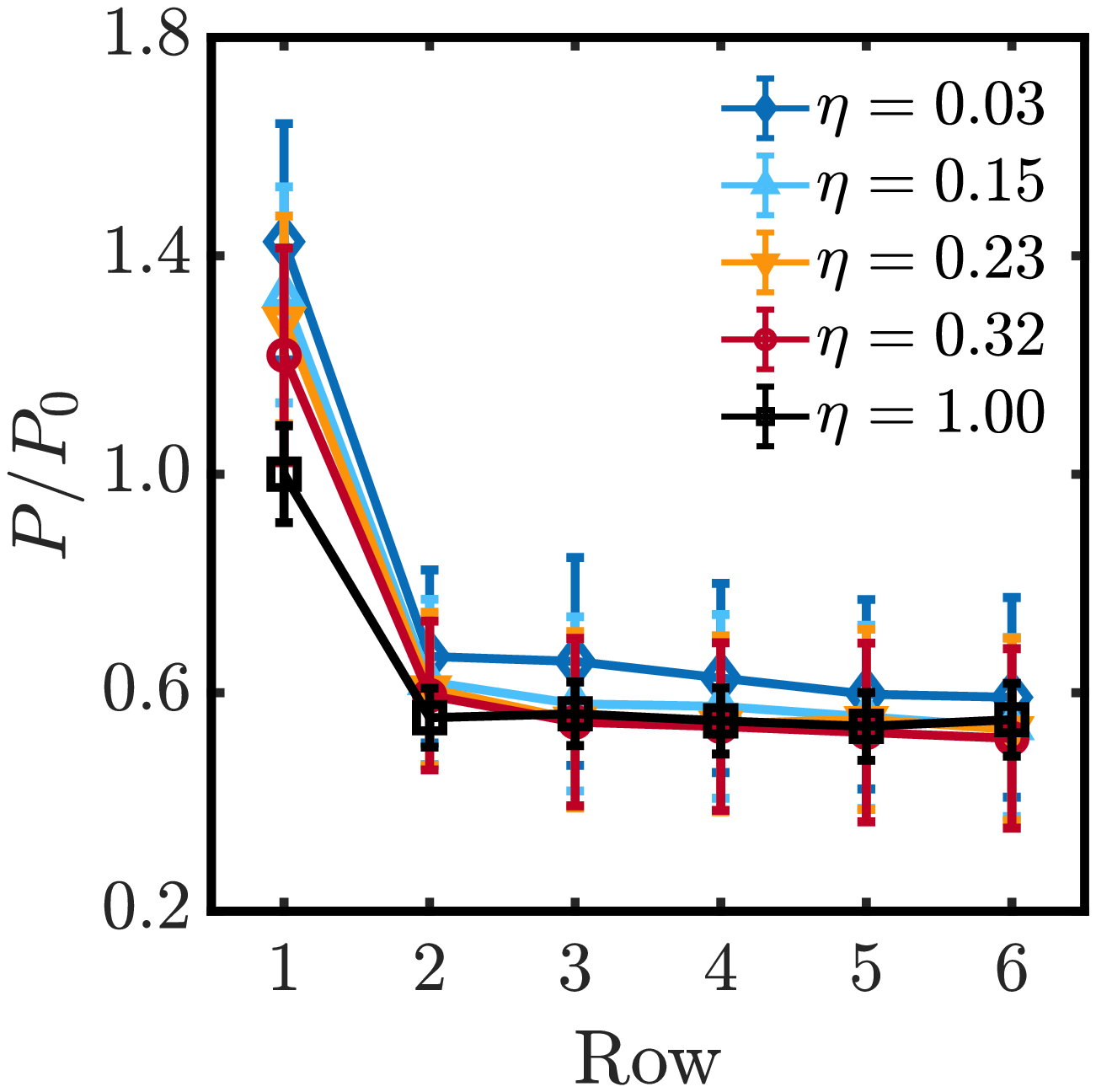} 
\put(0,88){$(b)$}
\end{overpic}
\begin{overpic}[width=0.3\textwidth]{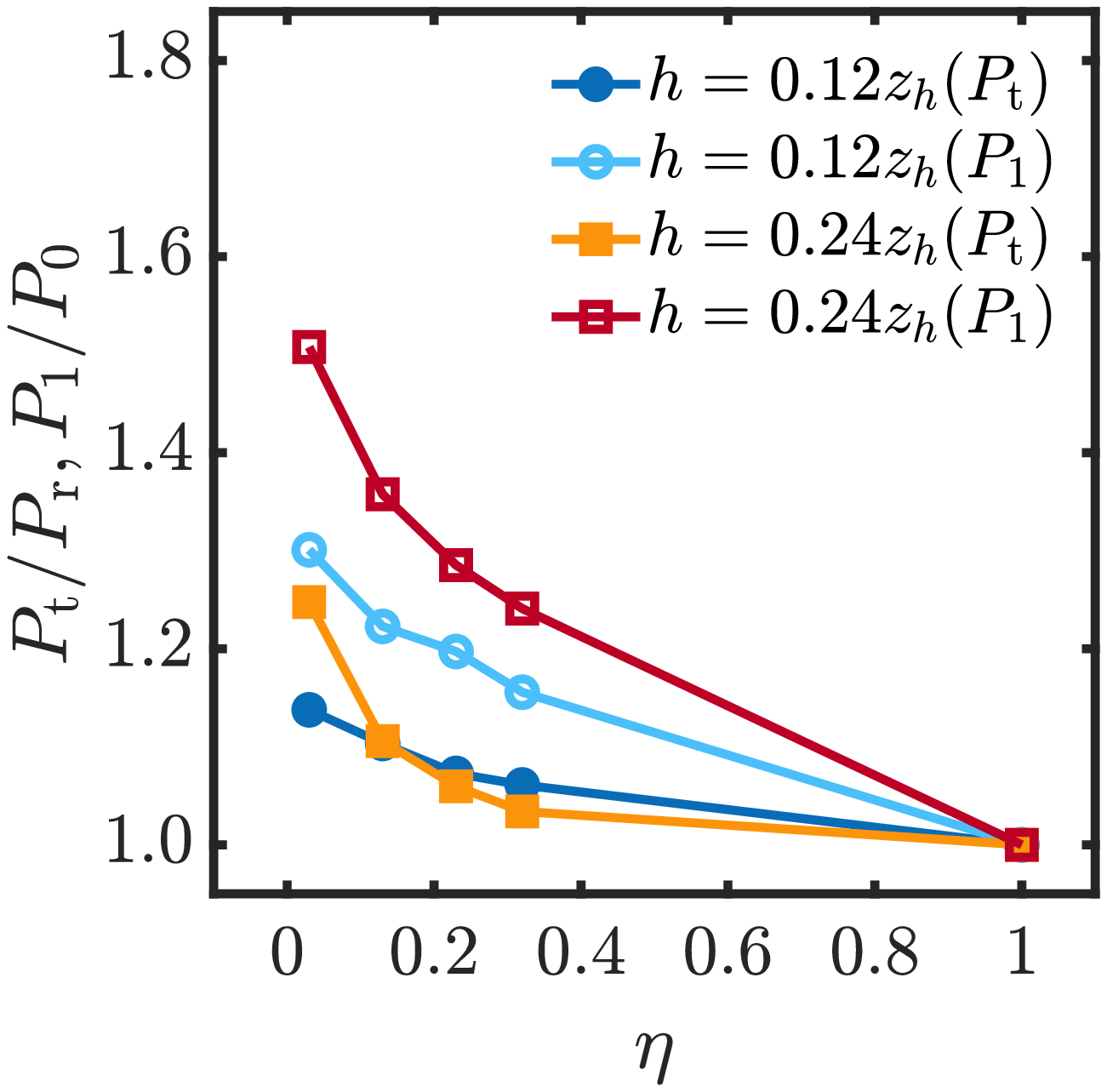} 
\put(0,88){$(c)$}
\end{overpic} 
\caption{(a,b) The normalized wind turbine power production $P/P_0$ as a function of downwind position for different windbreak porosity when (a) $x_t/h=6, h/z_h=0.12$ and (b) $x_t/h=2, h/z_h=0.24$. (c) The corresponding normalized wind farm ($P_{\rm t}/P_{\rm r}$, filled symbols) and first row ($P_1/P_0$, open symbols) power production as a function of windbreak porosity. }
\label{fig.power-eta}
\end{figure}

\begin{figure} [!tb]
\centering
\begin{overpic}[width=0.3\textwidth]{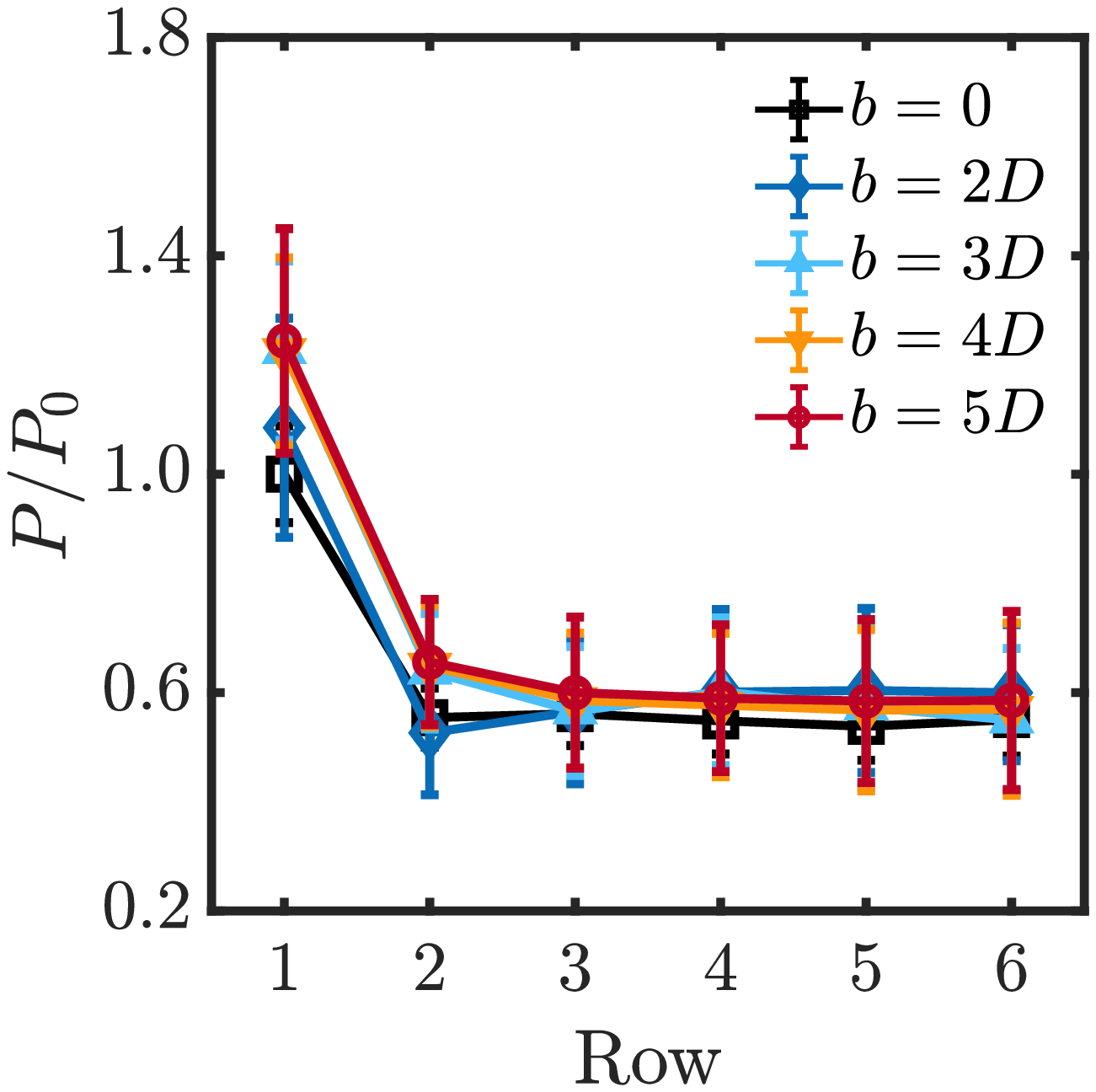}
\put(0,88){$(a)$}
\end{overpic} 
\begin{overpic}[width=0.3\textwidth]{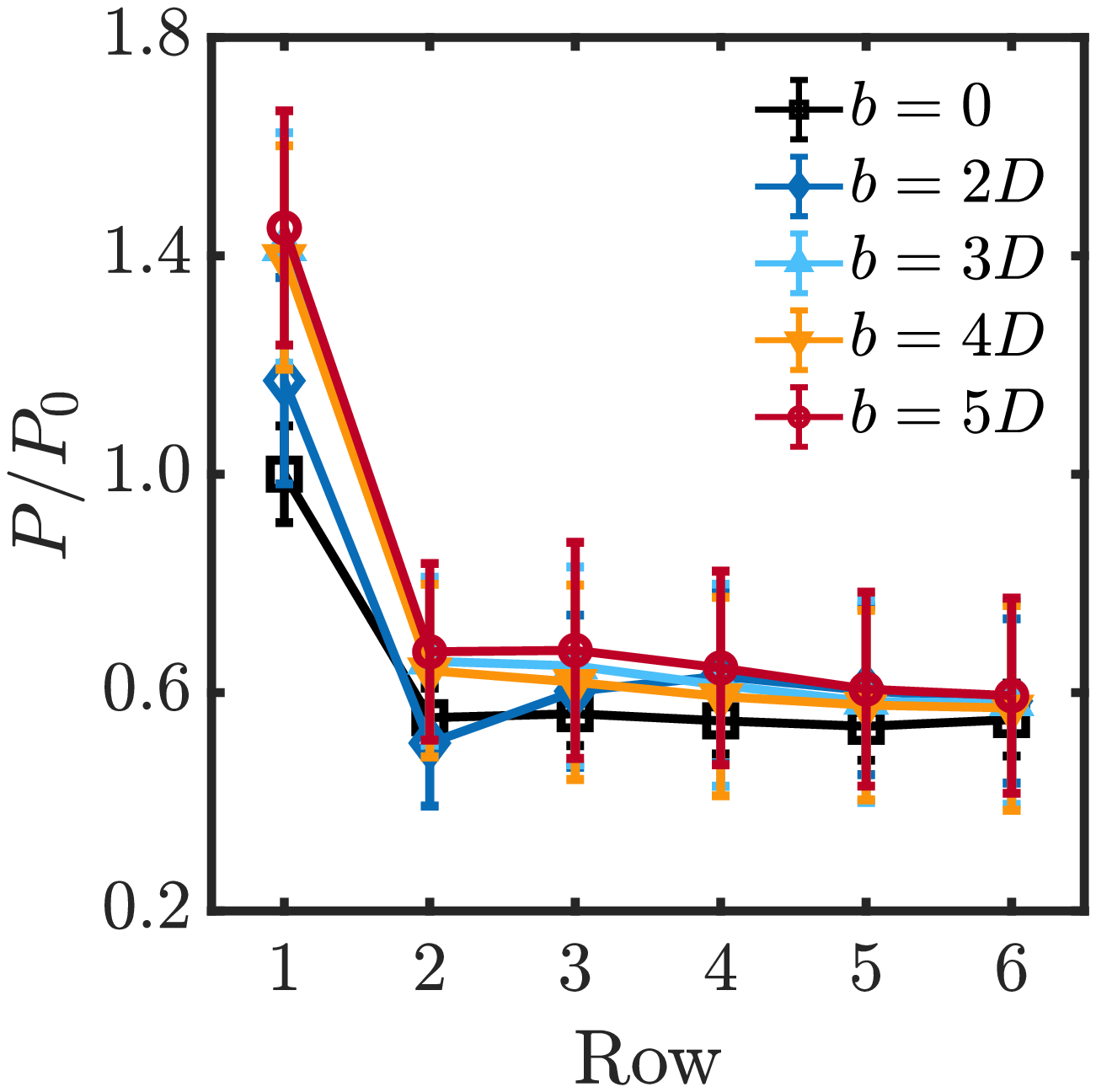} 
\put(0,88){$(b)$}
\end{overpic}
\begin{overpic}[width=0.3\textwidth]{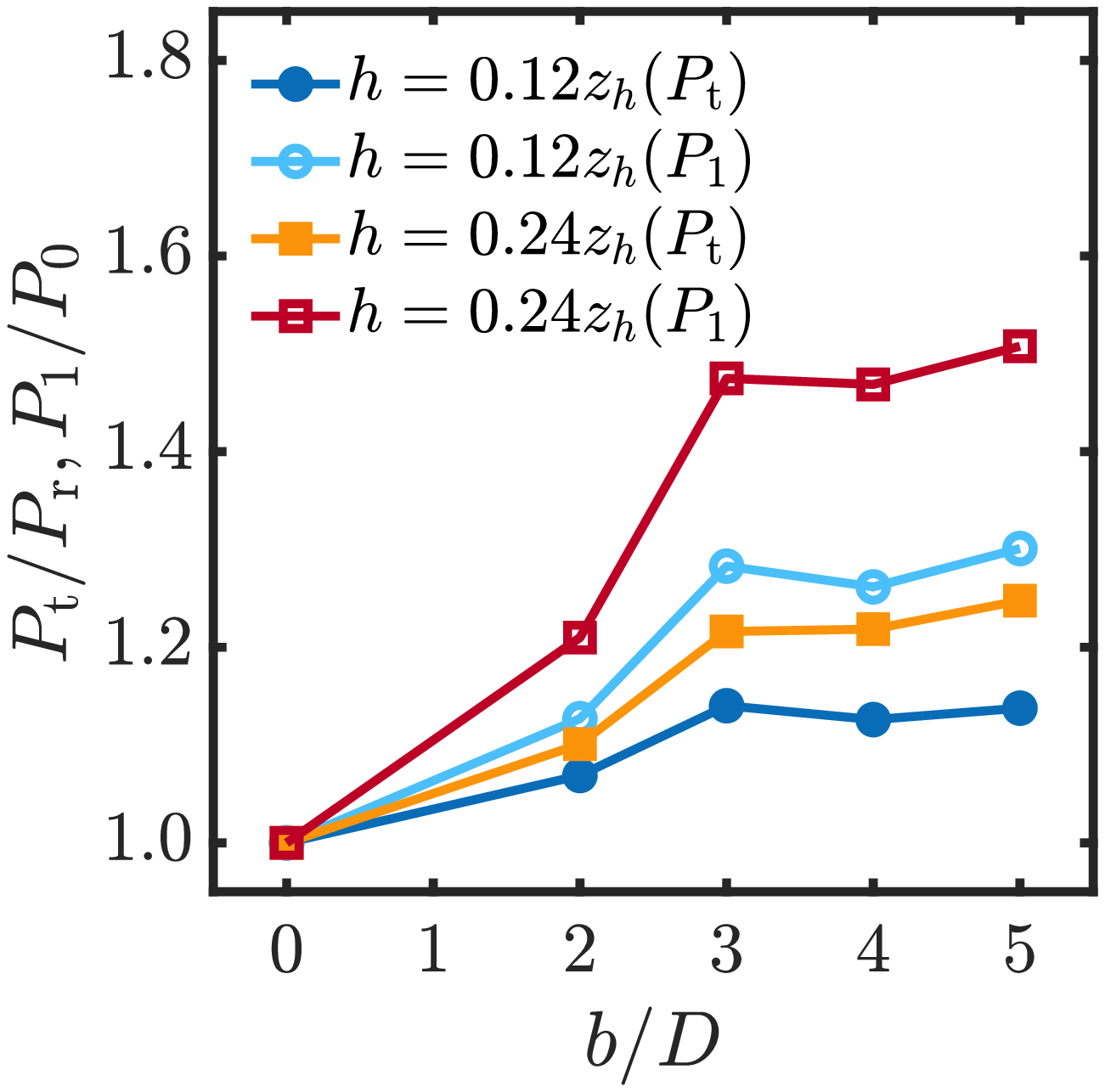} 
\put(0,88){$(c)$}
\end{overpic} 
\caption{(a,b) The normalized wind turbine power production $P/P_0$ as a function of downwind position for different windbreak widths $b$ when (a) $x_t/h=6$, $h/z_h=0.12$ and (b) $x_t/h=2$, $h/z_h=0.24$. (c) The corresponding normalized wind farm power production ($P_{\rm t}/P_{\rm r}$, filled symbols) and the first row power production ($P_1/P_0$, open symbols) as a function of windbreak width. The porosity is $\eta=0.03$.}
\label{fig.power-width}
\end{figure}

In Fig.~\ref{fig.power-eta} we investigate the effect of the windbreak porosity. Figure~\ref{fig.power-eta}(a,b) shows the normalized power production as a function of downwind position, and Fig.~\ref{fig.power-eta}(c) shows the wind farm production and the first-row production compared to the reference case without windbreaks. We find that windbreaks become more effective with decreasing porosity $\eta$, but this trend is much more pronounced for the first row than for the wind farm performance. The reason is that the first-row performance only depends on the speed-up effect obtained by the windbreak, which increases with decreasing $\eta$, while the wind farm performance is also affected by the increased drag imposed by the windbreaks. In Fig.~\ref{fig.power-width}, we investigate the effect of the horizontal extent of the windbreaks. The figure shows that increasing the windbreak width up to $3D$ has a positive effect on the performance of the wind farm. However, increasing the windbreak width beyond $3D$ does not seem to benefit the performance much.

\section{Conclusions}\label{sec.conclusions}

In agreement with experimental findings \citep{fan97, don07, tob17a} we show that windbreaks can increase wind turbine power production. In agreement with \citet{tob17a}, we find that the power production increases approximately linearly with windbreak height for low windbreaks. However, for higher windbreaks, the performance of downwind turbines is negatively affected by the windbreak wake. A crucial finding of the present study is that windbreaks can increase the power production of large wind farms. Previously it had been argued by \citet{tob17b} that the increased drag imposed by the windbreaks makes their use ineffective in very large wind farms. Indeed we find that the optimal windbreaks for a singular wind turbine can reduce wind farm production. However, a crucial finding of the present study is that lower windbreaks can significantly improve wind farm power production. A kinetic energy budget analysis reveals that the higher power production results from the favorable total pressure flux created. Besides, we show that the windbreaks effectiveness increases with decreasing porosity and saturates when the windbreak width is increased beyond three turbine diameters.

In addition to the parameters studied in this paper, we remark that many other parameters such as the inter-turbine spacing, wind farm layout, wind farm size, wind direction, thermal stratification, and complex terrain could significantly affect the optimal windbreak height. For example, the benefit of windbreaks and the optimal windbreak height depend on the wind farm size. In this study, we showed that the benefit of windbreaks is most prominent for an isolated row and smaller for a wind farm case with six rows, while \citet{tob17b} showed with simulation and model calculations that windbreaks become ineffective in infinite wind farms. Therefore, we hypothesize that the optimal windbreak height, and the corresponding increase in power production, decrease monotonically with increasing wind farm size. Similarly, the wind farm layout can be important as in a staggered wind farm, the inter turbine distance between consecutive turbines is larger than in an aligned wind farm layout, resulting in a higher optimal windbreak height in staggered farms than in aligned wind farms. However, we note that more extensive simulations are required to confirm these conjectures, and therefore, we leave this to future work.

\begin{acknowledgments}
We thank Leonardo Chamorro for insightful discussions. We thank Srinidhi N. Gadde for drawing the layout sketch in figure~\ref{fig.sketch} and making the video in the Supplemental Material. This work is part of the Shell-NWO/FOM-initiative Computational sciences for energy research of Shell and Chemical Sciences, Earth and Live Sciences, Physical Sciences, FOM, and STW and an STW VIDI grant (No. 14868). This work was partly carried out on the national e-infrastructure of SURFsara, a subsidiary of SURF corporation, the collaborative ICT organization for Dutch education and research. We acknowledge PRACE for awarding us access to MareNostrum 4 based in Spain at the Barcelona Computing Center (BSC) under Prace project 2018194742.
\end{acknowledgments}

\bibliography{literature_windfarms}

\end{document}